\pdfoutput=1
\documentclass[journal]{IEEEtran}

\usepackage{graphicx}
\usepackage{dcolumn}
\usepackage{bm}

\begin{document}
%
\title{Progress in atomic fountains at LNE-SYRTE}
%
%
%

\author{J.~Gu\'{e}na, 
        M.~Abgrall, 
        D.~Rovera, 
        P.~Laurent, 
        B.~Chupin, 
        M.~Lours, 
        G.~Santarelli, 
        P.~Rosenbusch, 
        M.E. Tobar, 
        R.~Li, 
        K.~Gibble, 
        A.~Clairon, 
        and S.~Bize
\thanks{Manuscript received September 15, 2011; accepted January 23, 2012.}
\thanks{This work was supported by  SYst\`{e}mes de  R\'{e}f\'{e}rence Temps-Espace
(SYRTE), the Laboratoire  National de M\'{e}trologie et d'Essais (LNE), the
Centre  National de la  Recherche Scientifique (CNRS), Universit\'{e} Pierre
et Marie curie (UPMC), the  Observatoire de Paris, the Institut Francilien de Recherche sur les Atomes Froids (IFRAF), the National Science
Foundation (NSF), Penn state, the Ville de Paris, the European Space
Agency (ESA), and the Centre  National d'Etudes Spatiales (CNES).}
\thanks{J. Gu\'{e}na, M. Abgrall, D. Rovera, P. Laurent, B. Chupin, M. Lours, G. Santarelli, P. Rosenbusch, A. Clairon and S. Bize
are affiliated to: LNE-SYRTE, Observatoire de Paris, CNRS, UPMC, 61 avenue de l'Observatoire, 75014 Paris, France}
\thanks{M. E. Tobar is affiliated to: The School of Physics, University of Western Australia, Crawley, Western Australia}
\thanks{R. Li and K. Gibble are affiliated to: The Department of Physics, The Pennsylvania State University, University Park, Pennsylvania 16802, USA}
\thanks{This work was supported by SYRTE, LNE,
CNRS, UPMC, Observatoire de Paris, IFRAF, NSF, Penn State, Ville de Paris, ESA and CNES}
\thanks{DOI: http://dx.doi.org/10.1109/TUFFC.2012.2208}
}

\maketitle

\begin{abstract}
We give an overview of the work done with the LNE-SYRTE fountain ensemble during the last five years. After a description of the clock ensemble, comprising three fountains FO1, FO2 and FOM, and its newest developments, we review recent studies of several systematic frequency shifts. This includes the distributed cavity phase shift which we evaluate for the FO1 and FOM fountains, applying the techniques of our recent work on FO2. We also report calculations of the microwave lensing frequency shift for the three fountains, review the status of the blackbody radiation shift, and summarize recent experimental work to control microwave leakage and spurious phase perturbations. We give current accuracy budgets. We also describe several applications in time and frequency metrology: fountain comparisons, calibrations of the international atomic time, secondary representation of the SI second based on the $^{87}$Rb hyperfine frequency, absolute measurements of optical frequencies, tests of the T2L2 satellite laser link, and review fundamental physics applications of the LNE-SYRTE fountain ensemble. Finally, we give a summary of the tests of the PHARAO cold atom space clock performed using the FOM transportable fountain.
\end{abstract}


%
\IEEEpeerreviewmaketitle

\section{Introduction}

Atomic fountain clocks provide the most accurate realization of the SI second and define the accuracy of the widely used international atomic time (TAI).
Ten years ago, a few formal TAI calibration reports were available each year. Now, there are normally several formal reports each month as a result of a great deal of
effort in several metrology institutes to improve the reliability and long term operation. At LNE-SYRTE, the fountain ensemble typically
provides more than 20 TAI calibrations each year, about half of all reports. These large improvements in reliability and corresponding capability to perform lengthy measurements yield an enhanced ability to investigate smaller and smaller frequency shifts, to compare fountain clocks with more stringent uncertainties and to test their long term behavior. In turn, applications using atomic fountain clocks have also developed and benefited from the improvements of the last few years.

In this article, we give an overview of developments and of applications of the LNE-SYRTE atomic fountain ensemble during the last five years.

\section{LNE-SYRTE fountain ensemble}

\subsection{Overview}

LNE-SYRTE has operated three atomic fountains for more than a decade. The first, FO1, is a $^{133}$Cs fountain which has been in operation since 1994 \cite{Clairon1995}. The second, FOM, was originally a prototype for the PHARAO\footnote{PHARAO
stands for ``Projet d'Horloge Atomique par Refroidissement d'Atomes en Orbite''} cold atom space clock \cite{Laurent1998} and was later modified to be a transportable fountain clock. The third, FO2, is a dual fountain which operates with $^{87}$Rb and $^{133}$Cs simultaneously \cite{Guena2010}. Development for more than a decade has continuously improved each of these. Our description here reflects the present configuration.

The three fountains share several features. All of them are gathering atoms in a Lin $\perp$ Lin optical molasses with laser beams oriented to launch atoms vertically in the so-called (1,1,1) configuration\footnote{In this configuration, the 3 pairs of counter-propagating laser beams for the molasses are aligned along the axes of a 3 dimensional orthonormal basis, where the (111) direction is along the vertical
direction, as sketched in Fig. \ref{fig_FontaineRbCs}} \cite{Gibble1993}. The 6 laser beams for the optical molasses are obtained from 6 fiber-coupled collimators, which are designed, machined and tuned to ensure proper alignment at the few $100~\mu$rad level when placed against the corresponding reference surfaces of the vacuum chamber. Similar fiber-coupled collimators provide the other laser beams for state selection and detection. In all three fountains, including the $^{87}$Rb part of FO2, the laser system is on a separate optical bench and the light is coupled to the vacuum system with optical fibers. The lasers are home-built extended cavity diode lasers, most of which use a narrow band interference filter for wavelength selection \cite{Baillard2006}.
This design is used in the PHARAO cold atom space clock (see \ref{subsec_PHARAO}). Also, it has recently become commercially available.
The required power is obtained either by injection locking another laser diode or using a tapered semi-conductor laser amplifier. In all three fountains, atoms are launched with the moving molasses technique \cite{Clairon1991} at a velocity of $\sim 4$~m.s$^{-1}$ and a temperature of $\sim 1~\mu$K. Atoms exit the launch sequence in the upper hyperfine state. For normal clock operation, microwave and laser pulses further select the atoms in the $m_F=0$ Zeeman sub-state of the lower hyperfine state ($|F=3,m_F=0\rangle$ for $^{133}$Cs, $|F=1,m_F=0\rangle$ for $^{87}$Rb).

All three fountains also have at least one layer of magnetic shield that surrounds the entire vacuum system. Notably, the molasses and the detection region are shielded, which is important to operate in the presence of the strong magnetic field fluctuations as of many urban environments, including the Observatoire de Paris. The largest, vertical magnetic field component in the lower part of the fountain (molasses and detection) is measured by a flux-gate magnetometer and actively stabilized with a set of horizontal coils distributed over the height of the fountain. Two or three additional cylindrical shields with end caps surround only the interrogation region to further attenuate magnetic field fluctuations. Inside the innermost shield, a solenoid and a set of compensation coils provide a static magnetic field in the range of $80$ to $200$~nT. The field is homogeneous to $10^{-3}$ and stable to $2\times 10^{-5}$ from 1~s to several days, as established by spectroscopy of the $|F,m_F=1\rangle \longrightarrow |F+1,m_F=1\rangle$ field sensitive transition. This corresponds to a stability of $\sim 4~$pT.

All three fountains also use the same strategy to control temperature and thermal gradients in the interrogation region. The innermost layer surrounding the interrogation is made of a material with high thermal conductivity (copper or aluminium alloy) and is well isolated from the environment to ensure good temperature uniformity of this layer. The temperature of this innermost layer is monitored using platinum resistors and is allowed to drift slowly, following temperature fluctuations in the room with a typical time constant of 1 day.
The measured temperature is used to infer the blackbody radiation temperature seen by the atoms. With this scheme, the temperature of the innermost layer is uniform to much better than $0.1$~K and no large gradients exist between the inner region and the environment. This essentially removes the need to care about the impact of the necessary openings (for the atomic trajectories, pumping,...) on the effective blackbody temperature for the atoms. Using heating devices to change or actively stabilize the temperature of the innermost layer induced a significant degradation of the temperature uniformity, which was difficult to compensate, even with segmented heaters. Consequently, this approach was abandoned.

\begin{figure}[h]
\includegraphics [width= \linewidth]{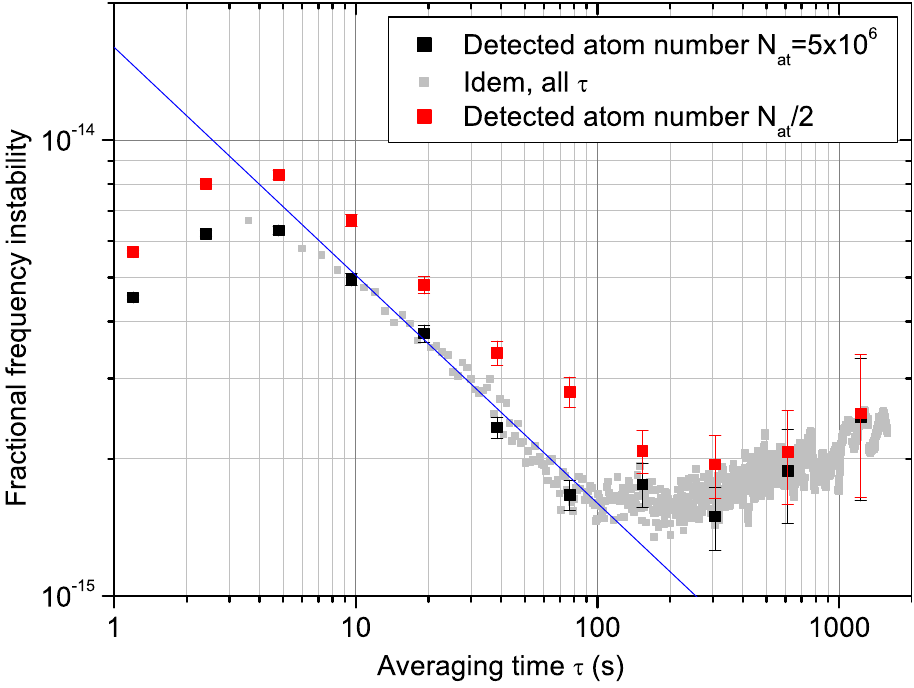}
\caption{Fractional frequency instability of FO2-Cs fountain when using a cryogenic sapphire oscillator as a reference signal. The black points are obtained for a high atom number: $N_{at}=5\times 10^{6}$ detected atoms. The red points are obtained for $N_{at}/2$ detected atoms. The blue line corresponds to $1.6\times 10^{-14}\tau^{-1/2}$. Above 100~s, frequency fluctuations of the ultra-stable interrogation signal (see \ref{subsec_ultrastablereference}), common to the high and low atom number measurements, are apparent.}\label{fig_beststability}
\end{figure}

The microwave interrogation cavities in all three fountains are $TE_{011}$ copper resonators with two independent microwave feedthroughs. The microwave synthesizers used to feed the cavities are home built. Despite significant differences, these synthesizers share the capability to use the ultra-low phase noise reference signal (see below \ref{subsec_ultrastablereference}) from a cryogenic sapphire oscillator (CSO) without significant degradation at the level of one to a few parts in $10^{15}$ at $1$~s \cite{Chambon2007}. Consequently, the short-term instability of all three fountains is typically limited by the atomic quantum projection noise \cite{Santarelli1999} and technical noise in the atom detection. The best short-term stability to date for the operation of an atomic fountain with a high atom number is $1.6\times 10^{-14}\tau^{-1/2}$ as reported in \cite{Vian2005} and exemplified in Fig. \ref{fig_beststability}. The present version of the synthesizers also incorporate a phase-stable microwave switch \cite{Santarelli2009} which helps to assess and remove leakage fields that may affect the interrogation process, independent of other systematic effects. Some details are given in \ref{subsec_microwaveleakage}.

For all three fountains, the computer system allows automated control of most of the important parameters, enabling routine implementation of multiple interleaved measurements with different configurations to study systematic shifts. A large number of parameters (laser powers, temperatures in the interrogation region, temperatures in the environment, at the Cs or Rb sources, humidity,...) are frequently monitored. The microwave power is regularly optimized and the magnetic field is verified via spectroscopy of a field sensitive transition, with automated computer controlled sequences.

\subsection{FO1}\label{subsec_FO1}

FO1 uses a 2 dimensional magneto-optical trap (2DMOT) \cite{Dieckmann1998} as a source of slow atoms to load the optical molasses. An advantage of this type of source is that it confines the high alkali vapor pressure to the 2DMOT chamber, yielding a large number of cold atoms in the molasses while keeping source consumption low. It also facilitates a better vacuum in the molasses region, thereby minimizing perturbations in the detection region and the contamination of the interrogation, including the microwave cavity, by background alkali vapor.
With higher laser powers in the 2DMOT, it should be possible to reproduce the atom numbers previously obtained with a chirped-slowed thermal atomic beam, that gave the best reported fountain stability \cite{Vian2005}, while consuming an order of magnitude less of alkali metal.
So far, the laser power devoted to the 2DMOT yields $\sim 4$ times fewer atoms in the molasses than with our best chirped-cooled atomic beam.
Note that 2DMOT loading of the molasses has some disadvantages.
Besides the quite intricated mechanical and optical design and the need for extra laser sources, we observe a degradation of the 2DMOT laser windows that are in contact with the relatively high alkali vapor pressure (up to $10^{-4}$~Pa) after a few years of almost continuous operation. Also, loading the molasses asymmetrically from the slow atom beam from the 2DMOT, leads to a non-spherical, complicated atomic cloud distribution which makes the investigation of some systematic frequency shifts more difficult. Note however, that by using the adiabatic passage method to change the atom number \cite{Pereira2002}, the extrapolation of the cold collision shift to zero density is essentially unaffected by distortions of the atomic cloud's shape.

The microwave cavity in FO1 has a diameter of $42.99$~mm and a length of $42.88$~mm. The below-cutoff guides that let the atoms pass through the cavity have a diameter of $12$~mm and a length of $60$~mm. At the two ends of the cutoff guides, an additional diaphragm with a diameter of $11$~mm blocks atoms that could travel close to the copper walls. Each of the 2 opposing microwave feedthroughs includes a resonant, polarization-filtering rectangular waveguide section. In the latest implementation, the cavity has a loaded quality factor of $14000$.

The interrogation region in FO1 is currently equipped with Stark plates. These can be used to measure the sensitivity of the clock transition to static electric fields to determine the Stark coefficient for the blackbody radiation shift correction. Alternatively, FO1 can be equipped with a blackbody radiator to directly measure the blackbody radiation shift (see \ref{subsec_BBR} below).

\subsection{FO2 dual Rb/Cs fountain}

\begin{figure}[h]
\includegraphics [width= 0.9\linewidth]{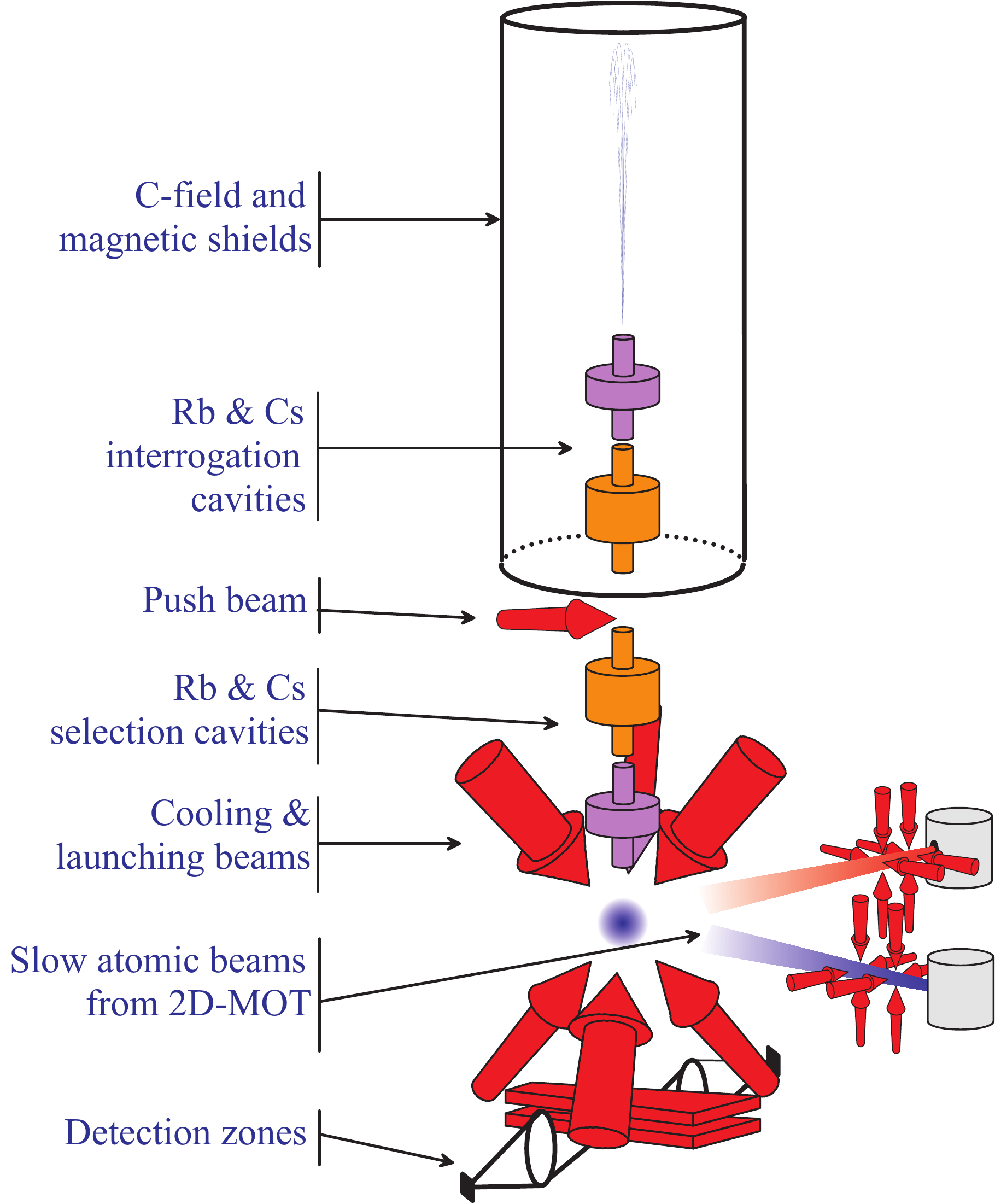}
\caption{Schematic view of the FO2 dual Rb/Cs fountain. The 852~nm light and the 780~nm light for $^{133}$Cs and $^{87}$Rb atoms are superimposed in each of the 9 laser beams with dichroic beam splitters. $^{133}$Cs and $^{87}$Rb atoms are captured, launched, state selected, probed and detected simultaneously at each fountain cycle of $1.6$~s. There are 2 independent 2D magneto-optic trap for the two atomic species.}\label{fig_FontaineRbCs}
\end{figure}

A schematic view of the FO2 dual Rb/Cs fountain is shown in Fig. \ref{fig_FontaineRbCs}. It uses 2 independent 2DMOT sources for Cs and for Rb. For Rb, only the $27.8\%$ abundant isotope is used in the fountain and the use of a 2DMOT eliminates the $^{87}$Rb background atoms, as for Cs, and also the larger background from the unwanted $^{85}$Rb isotope. The Rb and Cs molasses are superimposed with dichroic collimators for each molasses beam, where the 852~nm light (Cs) and the 780~nm light (Rb) come from 2 independent optical benches through polarizing fibers and are then combined with a dielectric dichroic beamsplitter and collimated with an achromatic lens \cite{Chapelet2008}. Similar dichroic collimators are used for all other laser beams, the pushing beam for the state selection, detection beams and repumper for detection.

As shown in Fig. \ref{fig_FontaineRbCs}, there are a total of 4 microwave cavities for state selection and interrogation of Rb and Cs atoms. Because both interrogation cavities are in the interrogation region, their temperature must be uniform and therefore the same as the environment temperature, which removes any possibility to independently tune each cavity. Practically, the Rb and Cs cavities are a single copper assembly where the Rb and Cs cavity resonances were mechanically tuned relative to one another, after a lengthy and tedious process. The resonance frequencies had to be verified under vacuum and tightly controlled thermal conditions, after each tuning step that was a controlled polishing of the two cavity spacers \cite{Sortais2001a}\cite{Bize2001a}. The Rb and Cs microwave cavities have the same aspect ratio, with the radii scaled by the ratio of the hyperfine frequencies. The cavity for Cs is located at the top. It has a diameter of $50.00$~mm and a length of $26.87$~mm. The Rb cavity has a diameter of $67.25$~mm and a length of $36.01$~mm. The below-cutoff end cap tubes have a diameter of $12$~mm and a length of $\sim 50$~mm. The 2 opposing microwave feedthroughs for each cavity include a resonant, polarization filtering rectangular waveguide section. The loaded quality factor is $7000$ for Cs and $6000$ for Rb.

The Rb and Cs clocks in FO2 routinely operate simultaneously since 2009 \cite{Guena2010}. Two independent but connected and synchronized computer systems allow the coordination of multiple interleaved measurements with Cs and Rb, as well as all other tasks for each clock, such as monitoring the temperature, the magnetic field, etc. Thus far, the nominal configuration is to launch the Rb and Cs clouds almost simultaneously, but at different velocities, so that the interrogation times are simultaneous, i.e. the Rb cloud is at the center of the Rb cavity when the Cs cloud is at the center of the Cs cavity, on both the way up and the way down. In this way, the two atomic clouds do not interact with each other during the interrogation process, avoiding interspecies collisions. The two clouds also remain well separated as they fall to the detection region, allowing the time-resolved detection of Rb and Cs \cite{Guena2010}. As a result, the Rb and Cs clocks  run simultaneously without impacting the performance of the other. The flexible computer systems allow for a large number of other possibilities which will be explored in the future. For instance, one cloud can be purposely launched through the other with a well controlled and yet variable speed, in order to study interspecies collisions at varying energies \cite{Legere1998}, based either on measurements of the cold collision frequency shift or on direct detection of the scattered wave \cite{Hart2007}.

\subsection{FOM transportable fountain}\label{subsec_FOM}

\begin{figure}[h]
\includegraphics [width= \linewidth]{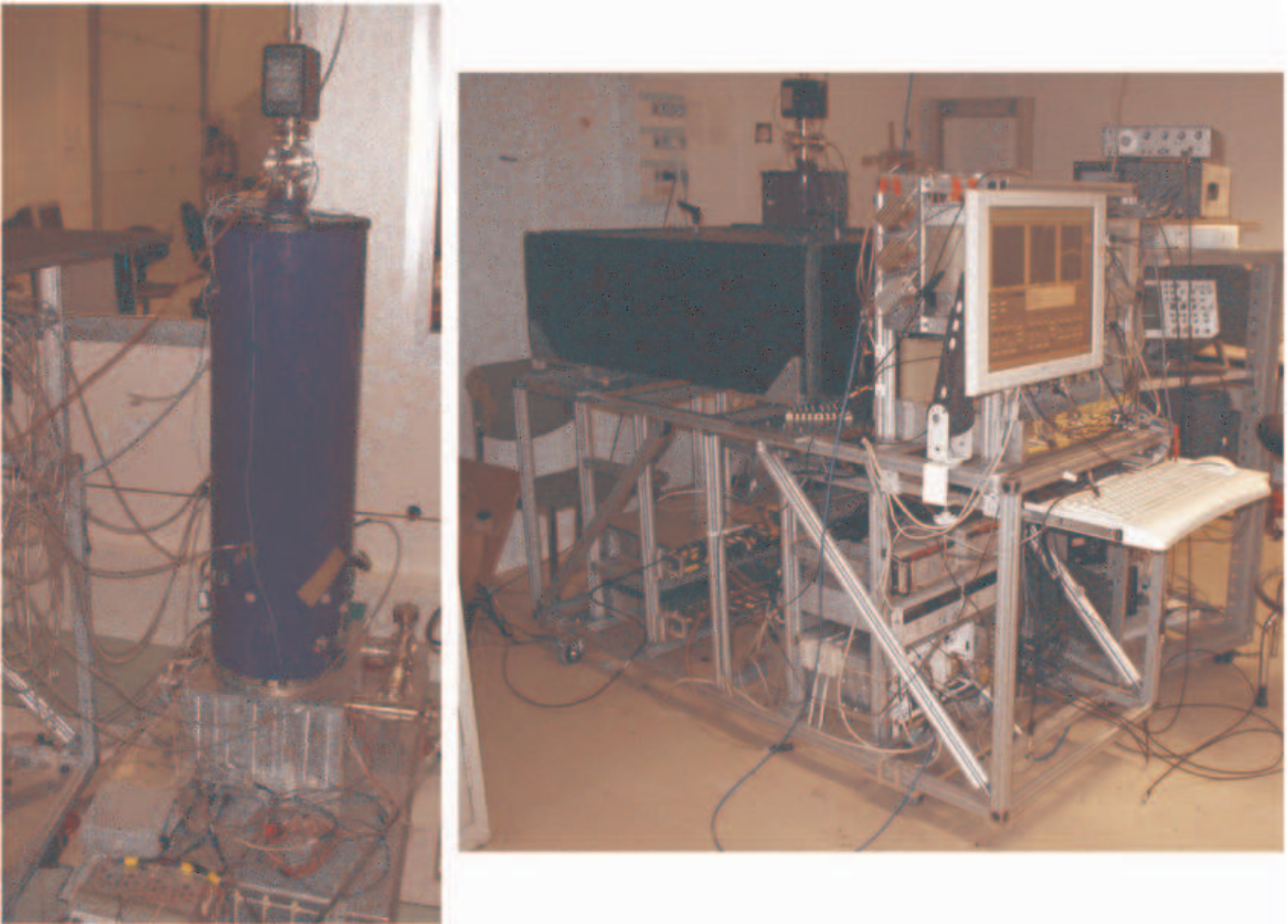}
\caption{Pictures of the transportable fountain FOM. On the left is the vacuum chamber for cooling and launching cesium atoms. Four layers of magnetic shielding surround the chamber. On the right is the second clock component with on top the optical bench inside an acoustic box, and below the electronics, including the microwave synthesis chain. A computer controls the clock.\label{fig_FOM}}
\end{figure}

Photographs of the transportable fountain FOM are shown in Fig. \ref{fig_FOM}. FOM has two major subsystems. The first is the vacuum system where atoms are manipulated. The second is the optical laser bench, its electronics, the microwave synthesizer and the computer system that operates the fountain, all of which are mounted on a single frame. Both of these subsystems include vibration damping devices necessary from transportation. When disconnected, the two subsystems fit into a small size truck. A battery powers the ion pump to maintain the vacuum of the system. After transportation, the two subsystems are reconnected (polarizing optical fibers for the laser light, microwave cables, wires for monitoring) and the system is allowed to thermally stabilize for a few hours. After a limited number of
manual re-optimizations (laser power through the optical fibers), a computer controlled sequence is started which automatically finds the laser set points and activates the laser stabilization loops \cite{Allard2004}. At remote locations, close-to-nominal operation of the fountain is typically recovered in less than 2 days.

One notable feature of the transportable fountain FOM is that the microwave synthesizer has two possible modes of operation. At LNE-SYRTE and at laboratories where a high quality metrological reference signal is available, the microwave synthesizer is synchronized to this external reference and the FOM fountain works as a frequency standard calibrating the frequency of the external reference. In the second configuration, FOM autonomously delivers a primary reference signal. Here, a BVA quartz oscillator synchronizes the microwave synthesizer and a digital loop locks the frequency of this BVA quartz oscillator to the atoms.

The microwave interrogation cavity in FOM has a diameter of $41.06$~mm and a length of $65.66$~mm. The two opposing microwave feedthroughs have direct evanescent coupling: the coaxial input cable is terminated in a small non-resonant cavity, with an antenna located a few millimeters away from the coupling iris of the cavity.
The cavity has a loaded quality factor of $21000$.

\subsection{Ultra-stable reference and frequency distribution scheme}\label{subsec_ultrastablereference}

\begin{figure}[h]
\vspace{-10mm}
\includegraphics [width= \linewidth]{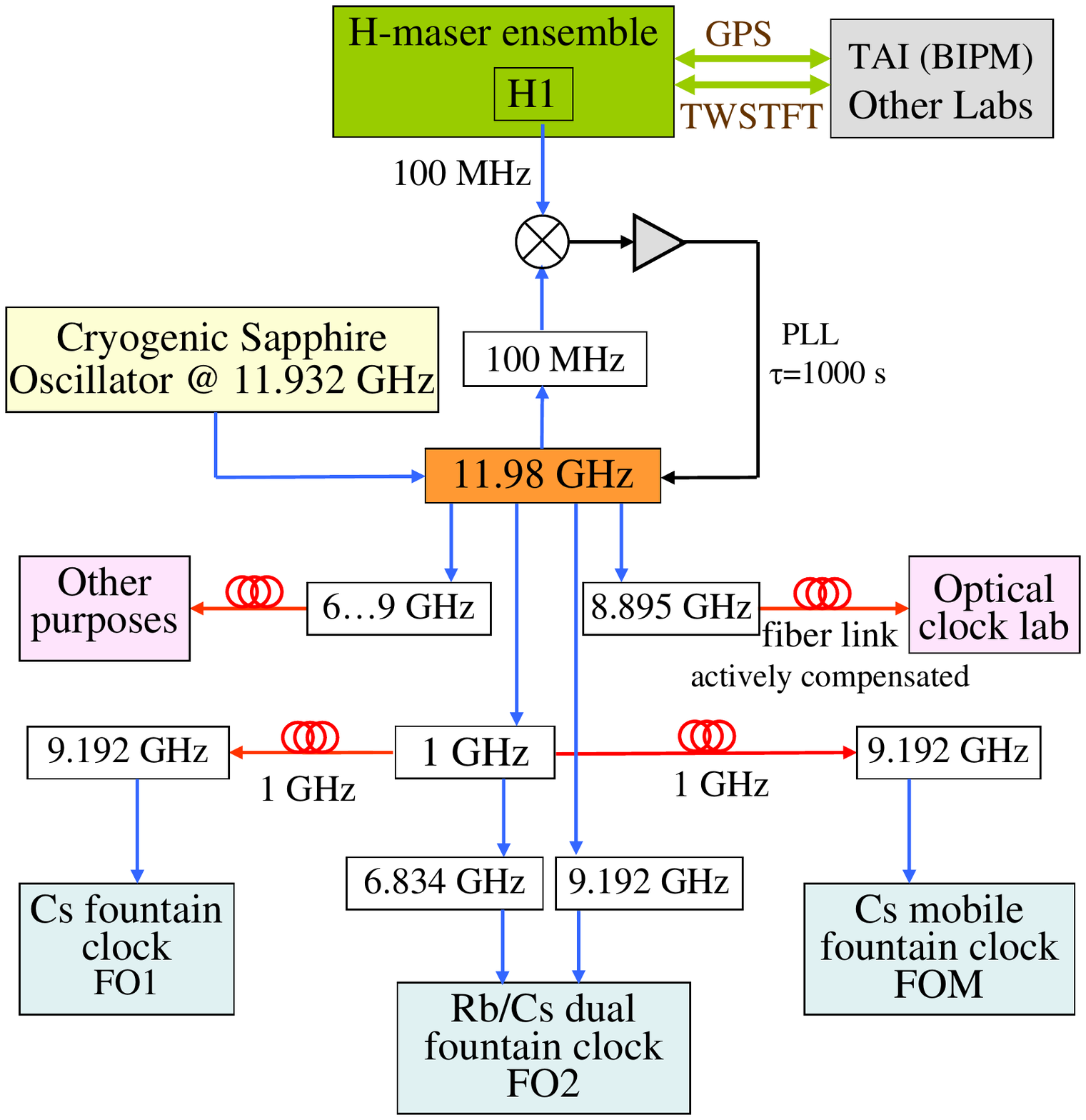}
\vspace{-40mm}
\caption{LNE-SYRTE clock ensemble. A Cryogenic Sapphire Oscillator generates an ultra-low noise reference signal that is distributed to
all the three fountains. This 11.98~GHz reference signal is slowly phase-locked (PLL) to the output of a hydrogen maser (H1), which provides long term stability and the connection to local timescales, satellite time transfer systems (GPS, TWSTFT), and the international atomic time (TAI). The reference signal can also be distributed through actively compensated fiber links.}\label{fig_ClockEnsemble}
\end{figure}

Figure \ref{fig_ClockEnsemble} is a schematic of the LNE-SYRTE fountain ensemble. A Cryogenic Sapphire Oscillator (CSO) \cite{Luiten1995b}\cite{Mann2001} is the local oscillator for all of the atomic fountains. Its ultra-low phase noise allows the fountains to operate at the quantum projection noise limit \cite{Santarelli1999}, producing short-term instabilities as low as $1.6\times 10^{-14}$ at 1~s (\cite{Vian2005} and Fig. \ref{fig_beststability}). As shown in Fig. \ref{fig_ClockEnsemble}, a finely tunable\footnote{with a computer controlled Direct Digital Synthesizer} microwave $11.98$~GHz signal is first derived from the free-running CSO. This frequency is divided down to 100~MHz and compared to the output of one of the LNE-SYRTE hydrogen masers, H1 in Fig. \ref{fig_ClockEnsemble}. A computer driven digital phase-lock loop (time constant $\sim 1000$~s) corrects the frequency of the $11.98$~GHz microwave signal to phase lock the 100~MHz signals.
This $11.98$~GHz reference has the exquisite phase noise properties of the CSO for time scales less than $\sim 1000$~s and the mid and long-term frequency stability of the hydrogen maser, while providing the connection to LNE-SYRTE timescales, to the International Atomic Time (TAI) and to other laboratories via GPS and TWSTFT satellite time transfer. This $11.98$~GHz signal is the high-performance reference from which all other signals are derived. A 1~GHz source is the starting point of the microwave synthesis for FO1, FO2-Rb and FOM. Since FO2 is located next door to the CSO, a more direct synthesis from 11.98~GHz to 9.192~GHz could be used for FO2-Cs \cite{Chambon2005}. To compare to optical clocks through optical frequency combs, a $8.985$~GHz is also generated. Finally, an additional synthesizer generates  signals in the 6 to 9~GHz range for testing (see \ref{subsec_microwaveleakage}). The design and characterization of the low phase noise electronics that generates these reference signals are reported in \cite{Chambon2005}\cite{Chambon2007}. Phase noise power spectral densities can be found in \cite{Mann2001} for the CSO alone, and in \cite{Chambon2005}, \cite{Chambon2007} and \cite{Dawkins2010} for the derived reference signals.

Figure \ref{fig_ClockEnsemble} shows 100 to 300~m long actively compensated optical fibers links that are used to distribute the reference signals. These links are simplified versions of similar long-distance links \cite{Daussy2005}, due to the short distance and limited insertion losses. The reference signal modulates the current of a $1.55~\mu$m diode laser that is launched into a standard telecom fiber. At the remote end, a fast photodiode detects
the intensity modulation and delivers the signal to the application. A fraction of this recovered signal is used to modulate a second diode laser that is launched back into the fiber. At the emission site, the returned signal is detected by another fast photodiode and then mixed with the reference signal to detect the phase variations imposed by the two passes propagation through the fiber. A temperature controlled fiber spool of 50 to 100~m is inserted to control the length of the link to cancel these phase variations with a computer operated servo-loop (time constant $\sim 10$~s). For the high performance links at $6-9$~GHz and $8.985$~GHz, an additional fast actuated piezo-driven fiber stretcher is used (bandwidth $\sim 400$~Hz). Characterization of these links has shown stabilities $\sim 10^{-14}$ at 1~s for the 1~GHz carrier and $\sim 2\times 10^{-15}$ at 1~s for the $6-9$~GHz and $8.985$~GHz carriers, and a residual frequency offset well below $10^{-16}$. A computer acquires and analyzes the doubled-passed phase signals to assess the performance in real-time and detect anomalies.

\subsection{New solution for the ultra-low noise reference oscillator}

The CSO allows a large improvement in the fountain short-term stability but requires cumbersome and costly liquid helium refills every 26 days.
Each refill lasts several hours and degrades the performance of the CSO for nearly half a day.
An alternate solution for fully continuous operation of an ultra-low noise microwave reference would be to use a pulsed-tube cryocooler  for the CSO
\cite{Hartnett2010a}\cite{Hartnett2010}\cite{Grop2010}\cite{Grop2009}.
However, advances in laser frequency
stabilization and optical frequency combs offer the possibility to replace the CSO with a fully continuous, cryogen-free and maintenance-free ultra-low noise
microwave generation system.

\begin{figure}[t]
\includegraphics [width= 0.95\linewidth]{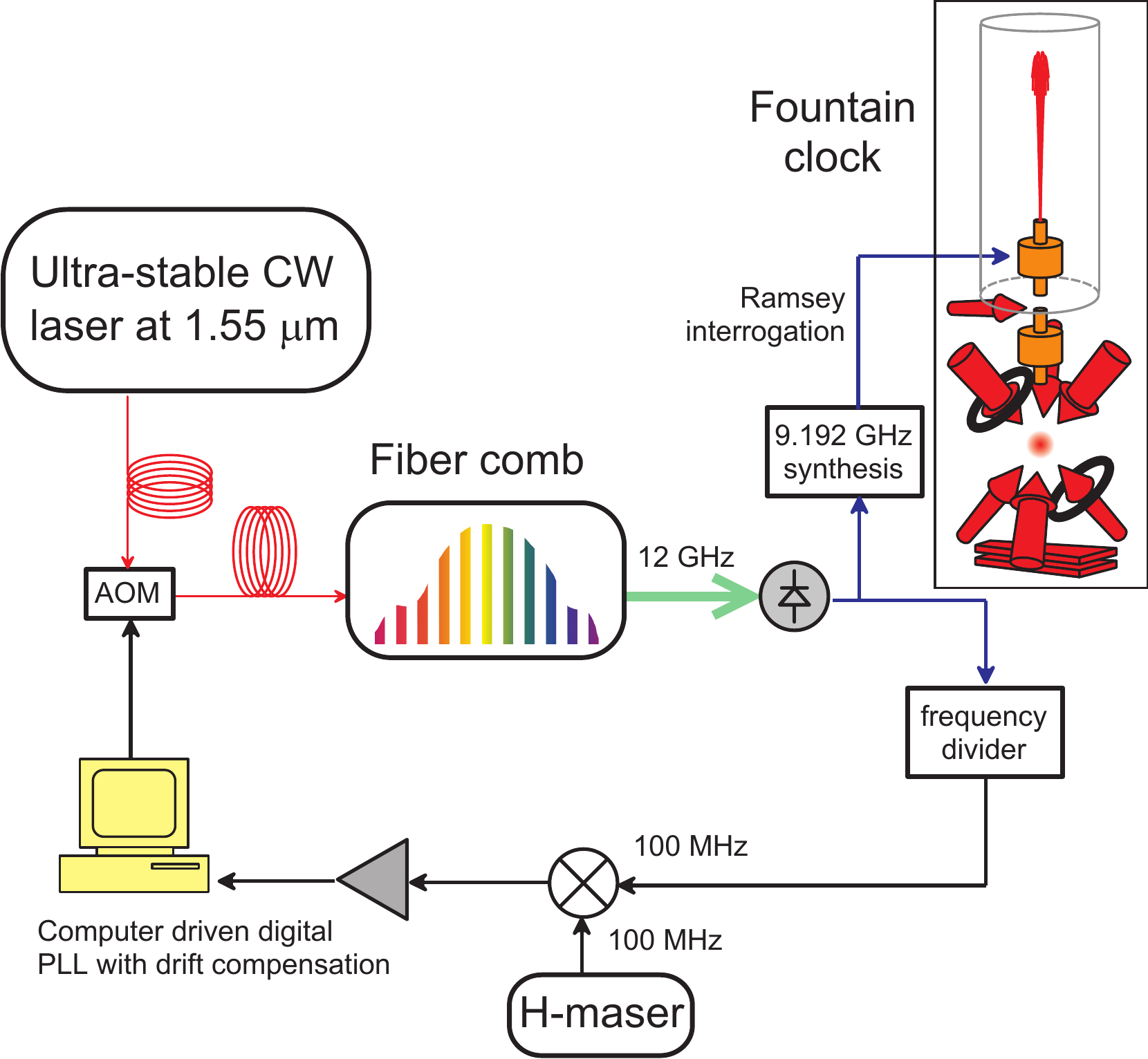}
\caption{An ultra-stable laser and an optical frequency comb produce an ultra-low noise microwave flywheel for the atomic fountains. The 12~GHz microwave signal is obtained by detecting a harmonic of the repetition rate of a Er-doped fiber femtosecond laser, which is stabilized to a CW laser itself stabilized to an ultra-stable Fabry-Perot cavity.\label{fig_FemtoComb}}
\end{figure}

Figure \ref{fig_FemtoComb} shows a schematic of such a laser-based system. An ultra-stable laser is realized by stabilizing a reliable $1.55~\mu$m laser, such as
an Erbium-doped fiber laser or an extended cavity laser, to an ultra-stable reference Fabry-Perot cavity. Several Fabry-Perot stabilized lasers have demonstrated short-term
stabilities well below
$10^{-15}$ \cite{Young1999}\cite{Millo2009b}\cite{Jiang2011}, similar to stabilities reported for cryogenic sapphire oscillators \cite{Mann2001}. Note however that simplified versions of these cavities can be sufficient for application to atomic
fountains. Microwave signals with short-term frequency stabilities in the mid $10^{-15}$ range at 1~s
are indeed sufficient to achieve the quantum projection noise limit at $10^{-14}$ at 1~s. This instability is not easy to surpass for other reasons: it is difficulty
to increase the number of useful atoms per cycle, to cope with the tremendous cold collision shift that would result from more atoms, and to suppress certain  detection noise sources. A short ($\leq 10$~cm) reference Fabry-Perot cavity with mirrors made of Ultra Low Expansion glass with a comparatively simple temperature
stabilization system can deliver instabilities less than $2\times 10^{-15}$ between 1 and 1000~s, which is well suited to atomic fountains.

As shown in Fig. \ref{fig_FemtoComb}, the ultra-stable laser is used to stabilize the repetition rate of the optical frequency comb generated by a fiber based femtosecond laser. A well-chosen harmonic, here 12~GHz, is detected and divided down to 100~MHz to allow a phase
comparison with a hydrogen maser, similar to what is done for the CSO (see \ref{subsec_ultrastablereference}). Here, a digital phase lock loop shifts the frequency of
the acousto-optic modulator (AOM) and thereby the ultra-stable light that controls the comb repetition rate, ensuring the long-term stability of the 12~GHz reference that ultimately provides the $9.192$~GHz interrogation signal for the fountains.

Ultra-low noise microwave generation with fiber-based optical frequency combs was demonstrated in \cite{Lipphardt2008} and \cite{Millo2009c}. Atomic fountains using microwave signals derived from optical references reported fountain short-term stabilities of $3.5\times 10^{-14}\tau^{-1/2}$ \cite{Millo2009a} and $7.4\times 10^{-14}\tau^{-1/2}$ \cite{Weyers2009a}. For both, the fountain stability was quantum projection
noise limited and the microwave signal alone was compatible with a fountain stability $\leq 10^{-14}\tau^{-1/2}$. Subsequent progress on fiber-based optical
frequency combs will enable the implementation of a fully operational version of the system of Fig. \ref{fig_FemtoComb}.

Note that an alternative approach for achieving quantum noise limited stability without an ultra-low noise microwave sources is the continuous atomic fountain \cite{Guena2007}\cite{DiDomenico2011a}.

\section{Recent studies of systematic shifts}

\subsection{Distributed cavity phase shift}\label{subsec_DCP}

The distributed cavity phase (DCP) shift is a residual first order Doppler shift that occurs when the moving atoms interact with the field inside the microwave cavity that has a spatial phase variation. Up until recently, measurements have been unable to reveal this effect and to reasonably agree with calculations. On the one hand, this was because the corresponding frequency shifts can be quite small, the measured frequency shifts give only indirect information about the phase variations, this shift and its changes with fountain parameters can be intricated with other systematic effects. On the other hand, accurately calculating the actual phase distribution and its corresponding frequency shift was a challenging numerical problem.

In \cite{Li2004} and \cite{Li2010b}, a new theoretical approach was proposed to solve the problem of the cavity field computation.
The field, represented explicitly as the sum of a large standing wave $\mathbf{H}_0(\mathbf{r})$ and of a small field, $\mathbf{g}(\mathbf{r})=\sum\mathbf{g}_m(\rho,z)\cos(m\phi)$,
that describes the effects of wall losses and cavity feeds, is expressed as an azimuthal Fourier series. Only the lowest azimuthal terms are relevant $m=0$, $m=1$ and $m=2$.
We have recently published measurements of the DCP shift performed with $^{133}$Cs in FO2 \cite{Guena2011} where, for the first time, quantitative agreement between theory and experiment is found. This study led to a better characterization and a quantitative evaluation of the contributions to the uncertainty of the DCP shift. It highlighted the importance of controlling the fountain geometry, notably the effective tilt of the fountain and the detection inhomogeneity, to reduce the DCP shift uncertainty. Based on the experimentally validated model, new cavity designs were proposed \cite{Li2010b} to reduce the DCP shift and its uncertainty. These measurements lowered the DCP shift uncertainty for FO2-Cs clock from $3\times 10^{-16}$ to $8.4\times 10^{-17}$ and defined a set of measurements to establish the DCP uncertainty of a fountain, an approach adopted in \cite{Li2011}\cite{Weyers2012}. Here, we apply this approach to evaluate the DCP uncertainty of FO1, FOM and FO2-Rb.

The $m=1$ term corresponds to power flowing horizontally from one side of the cavity to the other. This contribution can be probed by measuring the fountain frequency as a function of the tilt of the launch direction. The uncertainty for this term is determined by the experimental uncertainty in frequency change as a function of tilt, after correctly balancing the feeds to null the tilt sensitivity, and by the uncertainty of the fountain tilt for nominal operation \cite{Guena2011}.
When we purposely exaggerate the $m=1$ DCP shifts by asymmetrically feeding the cavities of FO1 and FOM, the measurements show tilt sensitivities that are much smaller than those of FO2-Cs. This is largely due to the higher quality factor of their cavities and to their smaller cavity apertures.

Also, when the two feeds are adjusted to produce equal Rabi pulse areas, FO1 and FOM show no measured tilt sensitivities at the level of $(1.5\pm 1.7)\times 10^{-16}$~mrad$^{-1}$ and $(1.8 \pm 1.1)\times 10^{-16}$~mrad$^{-1}$ respectively, along the axis of the feeds. For FO2-Rb, this sensitivity is $(1.8\pm 1)\times 10^{-16}$~mrad$^{-1}$.

Tilt sensitivities in the perpendicular direction remain to be measured for FO1 and FOM. Based on the findings of \cite{Guena2011}, we can reasonably apply, for the time being, the same bound as in the parallel direction. For FO2-Rb, the measured perpendicular sensitivity is $\leq 0.9\times 10^{-16}$~mrad$^{-1}$.
Using the same methods as in \cite{Guena2011} (detected atoms \emph{versus} tilt, frequency difference between the 2 asymmetric feedings), we estimate the uncertainty on the effective launch direction to be $1~$mrad for FO1, $0.2~$mrad for FO2-Rb and $0.7$~mrad for FOM in the direction of the feeds and $0.7~$mrad, $0.7~$mrad and $0.7~$mrad in the direction perpendicular to the feeds.

\begin{figure}[h]
\includegraphics [width= \linewidth]{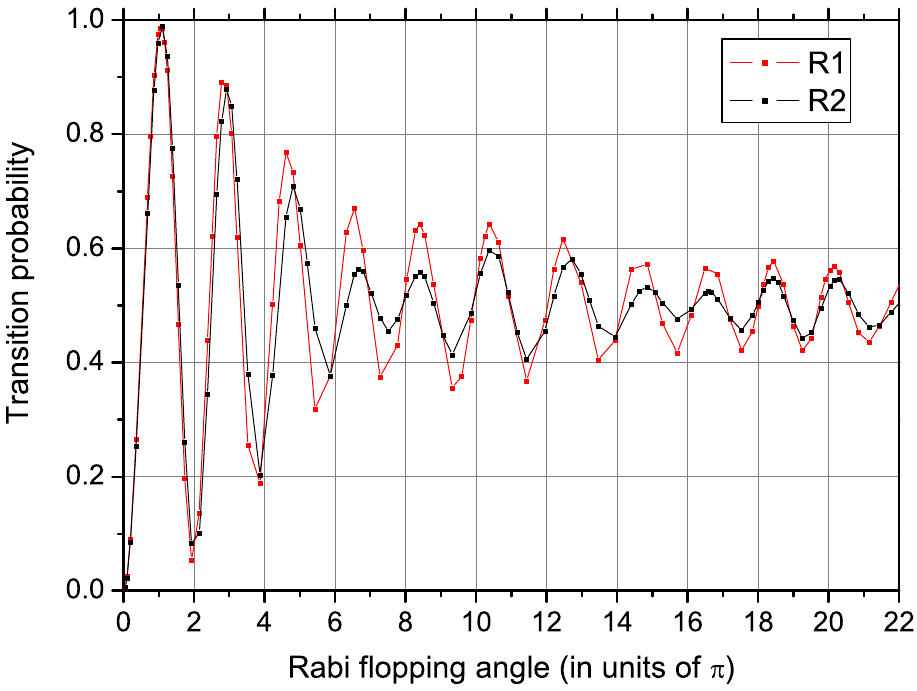}
\includegraphics [width= \linewidth]{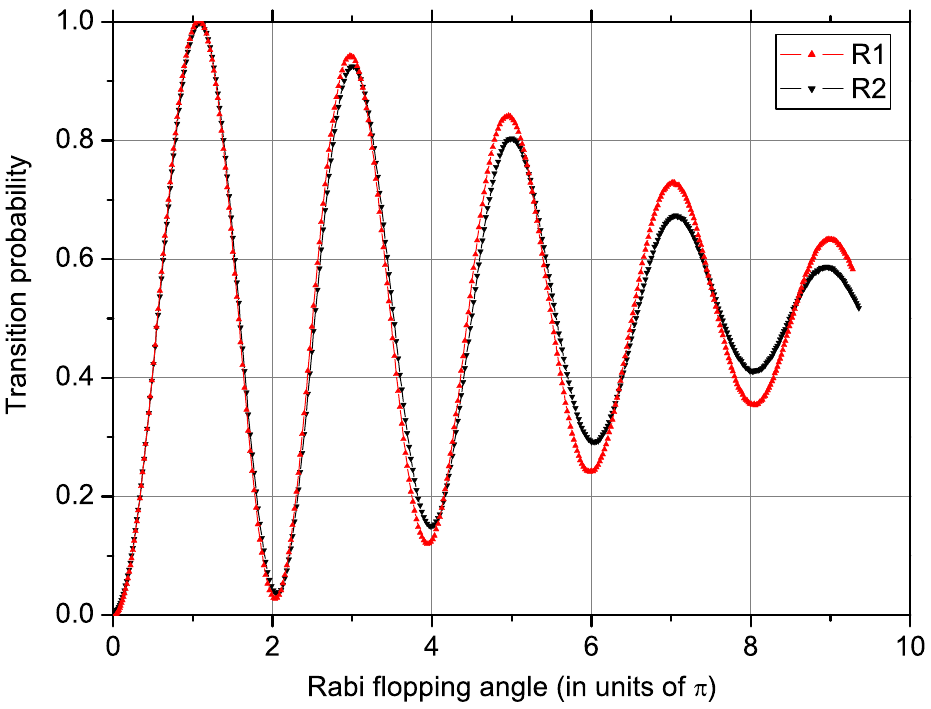}
\caption{Measured transition
probabilities for resonant Rabi pulses on the
upward (R1) and downward (R2) cavity traversals, as a function of the
microwave amplitude of FO1 (top graph) and FOM (bottom graph).\label{fig_contrastFO1FOM}}
\end{figure}

The $m=0$ and $m=2$ terms are calculated from the phase distributions \cite{Li2004}\cite{Li2010b} and the corresponding response of the fountain. This second step takes as an input a number of parameters of the fountain geometry such as the cloud size, the cloud temperature and the geometry of the detection. Figure \ref{fig_contrastFO1FOM} shows the measured contrast of the Rabi oscillations at upward and downward passages in the microwave cavity, which is used to ascertain the fountain geometry to compute the DCP shift \cite{Guena2011}. The same curve, measured for FO2-Cs, is shown in \cite{Guena2011a}.

The resulting DCP shift corrections and their uncertainties are presented in Table \ref{tab_DCP}. Several of these uncertainties are expected to decrease when improved measurements of the tilt sensitivities and tilt uncertainties will be available.

\begin{table*}
\renewcommand{\arraystretch}{1.3}
\begin{center}
\begin{tabular}{|c|c|c|c|c|c|}

\hline

$m$ & \small{FO1}  & \small{FO2-Cs}  & \small{FOM}  & \small{FO2-Rb} & \small{Source} \\

\hline

0 & -0.7   $\pm$ 0.4  & -0.4 $\pm$ 0.2  & -1.5  $\pm$ 0.7  & -0.2    $\pm$ 0.1 & \small{homogeneous wall losses} \\
  & 0      $\pm$ 0.8  & 0    $\pm$ 0.7  & 0     $\pm$ 1.2  & 0       $\pm$ 0.5 & \small{$20\%$ top/bottom loss asymmetry} \\
  & -0.7   $\pm$ 0.9  & -0.4 $\pm$ 0.8  & -1.5  $\pm$ 1.4  & -0.2    $\pm$ 0.5 & \small{total $m=0$} \\

\hline

1 & -9 $\pm$ 18 ($\dag$)& 0 $\pm$ 4.4 & 0 $\pm$ 13& 9.0 $\pm$ 6.2 ($\dag$) & \small{$\parallel$ to feeds} \\

  & 0 $\pm$ 22         & 0 $\pm$ 6.3 & 0 $\pm$ 13 & 0 $\pm$ 6.2 & \small{$\perp$} to feeds \\

  & -9 $\pm$ 27      & 0 $\pm$ 7.1 & 0 $\pm$ 16 & 9.0 $\pm$ 9.5 & \small{total $m=1$} \\

\hline

2 & 0 $\pm$ 1.4 ($\ddag$) & -8.8 $\pm$ 4.4 & -5.6 $\pm$ 2.8 & -5.1 $\pm$ 2.6 & \small{detection inhomogeneity} \\
  & 0 $\pm$ 3.2 & 0 $\pm$ 3.0 & 0 $\pm$ 2.2 & 0 $\pm$ 1.3 & \small{2~mm offset, homogeneous det.} \\
  & 0 $\pm$ 3.5 & -8.8 $\pm$ 5.3 & -5.6 $\pm$ 3.5 & -5.1 $\pm$ 2.9 & \small{total $m=2$} \\

\hline

 & -9.7 $\pm$ 27 & -9.2 $\pm$ 8.9 & -7.1 $\pm$ 16& 3.7 $\pm$ 9.9  & \small{total DCP} \\

\hline

\end{tabular}
\end{center}

\caption{Corrections and their uncertainties for the distributed cavity phase shift in FO1, FO2, and FOM, in units of $10^{-17}$. These corrections and uncertainties are given separately for each azimuthal term of the phase distribution, labeled with the azimuthal number $m$. The last row is the total DCP correction and uncertainty, given by the combination of $m=0,1~\&~2$.
$\dag$ For technical reasons not detailed here, FO1 and FO2-$\mathrm{Rb}$ are to date operating at a non-vanishing parallel effective tilt, hence the non-zero $m=1$ correction. $\ddag$ the first $m=2$ correction for FO1 is zero with a small uncertainty because of a $(45\pm 5)^{\circ}$ angle between the feed axis and the detection axes.} \label{tab_DCP}
\end{table*}

\subsection{Microwave lensing}

The external degrees of freedom of the atoms evolve dynamically during
the Ramsey interrogation. The interaction of the atoms with the
standing-wave field in the microwave cavity on the upward passage
slightly modifies the atomic motion, leading to a predicted shift of the
clock frequency. This effect was first called ``the microwave recoil"
\cite{Wolf2004}, connecting to the simple situation where an atom
interacts with a plane wave as in optical spectroscopy \cite{Hall1976}.
In the optical case, an atom absorbs a photon from a plane wave. The
shift, expressed in energy, is $(\hbar k )^2/ (2 m_{at})$, corresponding
to the kinetic energy of the atom with the momentum $\hbar k$ of the
absorbed photon. For microwave photons, the corresponding fractional
frequency shift is $1.5\times 10^{-16}$ for $^{133}$Cs, comparable to
the best reported fountain accuracy.

In 2006 a new approach to this effect provided better physical insight
and made the calculations tractable to treat both transverse dimensions
\cite{Gibble2006}. The microwave standing wave and its spatially varying
field amplitude produces a radial magnetic dipole force which acts as a
lens to focus (or defocus) the atomic wave packets, hence the name
``microwave lensing frequency shift". To calculate the shift with
reasonable accuracy, it is crucial to properly treat the apertures in the
fountain. They truncate the downwardly traveling atomic cloud and this
introduces a difference between the two dressed states, which experience
opposite lensing deflections on the upward cavity traversal.

Here, we report calculations of the microwave lensing for the LNE-SYRTE
fountains. We apply the approach that was recently used to
evaluate the NPL-CsF2 and PTB-CSF2 fountains
\cite{Li2011}\cite{Weyers2012}. The results are
summarized in Table
\ref{tab_MicrowaveLensing_a}. The simple ``analytic" result
\cite{Li2011} (1$^{\mathrm{st}}$ line in Table
\ref{tab_MicrowaveLensing_a}) neglects all apertures on the upward
passage and only treats dipole energy that is quadratic ($\propto \rho^2$) in the radial coordinate $\rho$. Since
the initial clouds in the LNE-SYRTE fountains are not point-like
sources, this overestimates the microwave lensing shift because
it includes atoms that would be cut by the cavity apertures, which experience much larger dipole forces away from
the cavity axis. Including an
aperture at the first cavity passage eliminates these atoms from the
calculation, giving a smaller and more realistic shift
(2$^{\mathrm{nd}}$ line in Table \ref{tab_MicrowaveLensing_a}). Adding
the lowest cutoff aperture for the upward passage, the variation of the Rabi
tipping angles, and the all-order dipole forces (beyond $\rho^2$), gives
a small correction (3$^{\mathrm{rd}}$ line), as does including the
detection inhomogeneities and imperfect Ramsey fringe contrast, yielding the predicted shift.

\begin{table*}
\renewcommand{\arraystretch}{1.3}
\begin{center}
\begin{tabular}{|l|c|c|c|c|}
\cline{2-5}
\multicolumn{1}{l|}{} &{\small FO1} &{\small FO2-Cs} & {\small FOM} & {\small FO2-Rb}\\
\cline{2-4}
\hline

{\small Analytic} & $7.8$ & $8.4$ & $11.9$ & $7.0$ \\
\hline
{\small Upward cavity aperture} & $6.1$ & $7.0$ & $9.2$ & $5.8$ \\
\hline
{\small All order dipole $\&$ upward cutoff aperture} & $5.7$ & $6.5$ & $8.6$ & $5.6$ \\
\hline
{\small Detection inhomogeneity $\&$ Ramsey contrast} & $6.5$ & $7.3$ & $9.0$ & $6.8$ \\


\hline
\end{tabular}
\end{center}

\caption{Calculated microwave lensing fractional frequency shifts for
FO1, FO2, and FOM, in units of $10^{-17}$.
The microwave lensing shift is given for each fountain with the
approximations described in the text. From top to bottom, the degree of sophistication of the model is
increasing. Our best final result is given in the last row.} \label{tab_MicrowaveLensing_a}
\end{table*}

\subsection{Blackbody radiation shift}\label{subsec_BBR}

The blackbody radiation (BBR) shift is one of the largest systematic corrections and a significant source of uncertainty in atomic fountains. The bulk of this shift is simply given by the scalar Stark shift of the clock transition \cite{Itano1982}, its response to a static electric field. A high accuracy measurement of the Stark coefficient was made early on at LNE-SYRTE with FO1 \cite{Simon1998} and subsequently used by most groups for their BBR corrections. These measurements where made with applied
electric fields ranging from $50$ to $150$~kV.m$^{-1}$, significantly higher than the $831.9$~V.m$^{-1}$ rms value of the BBR field at 300~K. The greatly improved
measurement capability enabled by the CSO (see \ref{subsec_ultrastablereference}) provides an opportunity to measure the Stark coefficient at an electric
field strength that is much closer to a typical BBR field.
The early measurements \cite{Simon1998} were specifically tested for a possible deviation from a purely quadratic
dependance on the electric field. This test did not show any such deviation, which was expected since there was no theoretical basis to expect significant higher order terms
given the measurement accuracy and field strengths.
Nevertheless, it was important to re-examine these measurements at much lower fields given the intervening conflicting measurements and calculations of \cite{Micalizio2004}\cite{Levi2004} and \cite{Godone2005}.

These new measurements \cite{Rosenbusch2007} at field strengths ranging from $1.5$ to $25$~kV.m$^{-1}$, are in excellent agreement with the
earlier measurements of \cite{Simon1998}.
To extract the scalar Stark coefficient for the BBR clock correction from these measurements, the small tensor contribution has to be subtracted \cite{Simon1998}.
In \cite{Ulzega2006}, a sign error in the earlier theory \cite{Sandars1967}\cite{Angel1968} for this tensor part was found, which was then confirmed experimentally
\cite{Ulzega2007}. Taking into account this revisited theory, we get a corrected value for the scalar Stark coefficient: $k_0=-2.282(4)\times 10^{-10}$~Hz.V$^{-2}$.m$^{2}$.
The change in the BBR fractional frequency correction from the previous value of \cite{Simon1998} is $6.8\times 10^{-17}$, $80\%$ of which
is due to the sign correction of the theory. The uncertainty of $k_0$ continues to be dominated by the uncertainty on the geometry of the copper plates used to apply the electric field in the fountain.
The value agrees with the most recent theoretical calculations \cite{Angstmann2006a}\cite{Beloy2006}\cite{Palchikov2003}.
It also agrees with several other calculations \cite{Feichtner1965}\cite{Lee1975}\cite{Itano1982}\cite{Micalizio2004},
after some \cite{Feichtner1965}\cite{Micalizio2004} are corrected for omitted contributions pointed out in \cite{Angstmann2006a} and \cite{Beloy2006}.
The only discrepancy that seems to remain, even after the omitted contributions are included, is the calculation of \cite{Ulzega2006} (see the table summary in \cite{Rosenbusch2007}). We also note that a thermal atomic beam experiment recently provided another Stark shift coefficient measurement \cite{Robyr2011}. The still preliminary value is in agreement with the LNE-SYRTE value and is $\sim 5$ times less accurate.

We have also performed direct measurements of the BBR shift by inserting a blackbody tube in the FO1 fountain \cite{Zhang2004}, exposing atoms to thermal radiation
at temperatures ranging from $300$ to $440$~K. Measurements were in excellent agreement with the above mentioned scalar Stark shift coefficient, but with a $22$ times larger uncertainty \cite{Rosenbusch2007}. On going comparisons between a room temperature and a cryogenic fountain
\cite{Heavner2011} should yield another measurement of the Cs BBR shift at room temperature.

Regarding the BBR coefficient of the $^{87}$Rb hyperfine frequency, the experimental status is unchanged from \cite{Bize1999}. On the
theoretical side, high accuracy \emph{ab initio} calculations have recently been published in \cite{Angstmann2006}\cite{Safronova2010a}.

\subsection{Microwave leakage}\label{subsec_microwaveleakage}

In the last few years, we have developed a new approach to assess frequency shifts due to microwave leakage. Ideally, after the state selection process, atoms interact
with the interrogating microwaves only in the microwave cavity during the upwards and downwards traversals. In practice, unintended interactions can occur
with a residual microwave field that is coherent with the probing field. Being unintended, such a residual field is by nature uncontrolled, with a potentially complex structure with both a
large standing-wave and a large traveling-wave component.
Such an extraneous field, which is difficult to exclude at the required level with purely
electric means, leads to a frequency shift which generally has a complicated variation with the fountain parameters, especially the microwave power
\cite{Weyers2006}\cite{Jefferts2005a}. If present, the frequency shift is generally difficult to disentangle from other systematic errors such as the distributed
cavity phase or the microwave lensing frequency shifts.

\begin{figure}[h]
\includegraphics [width= \linewidth]{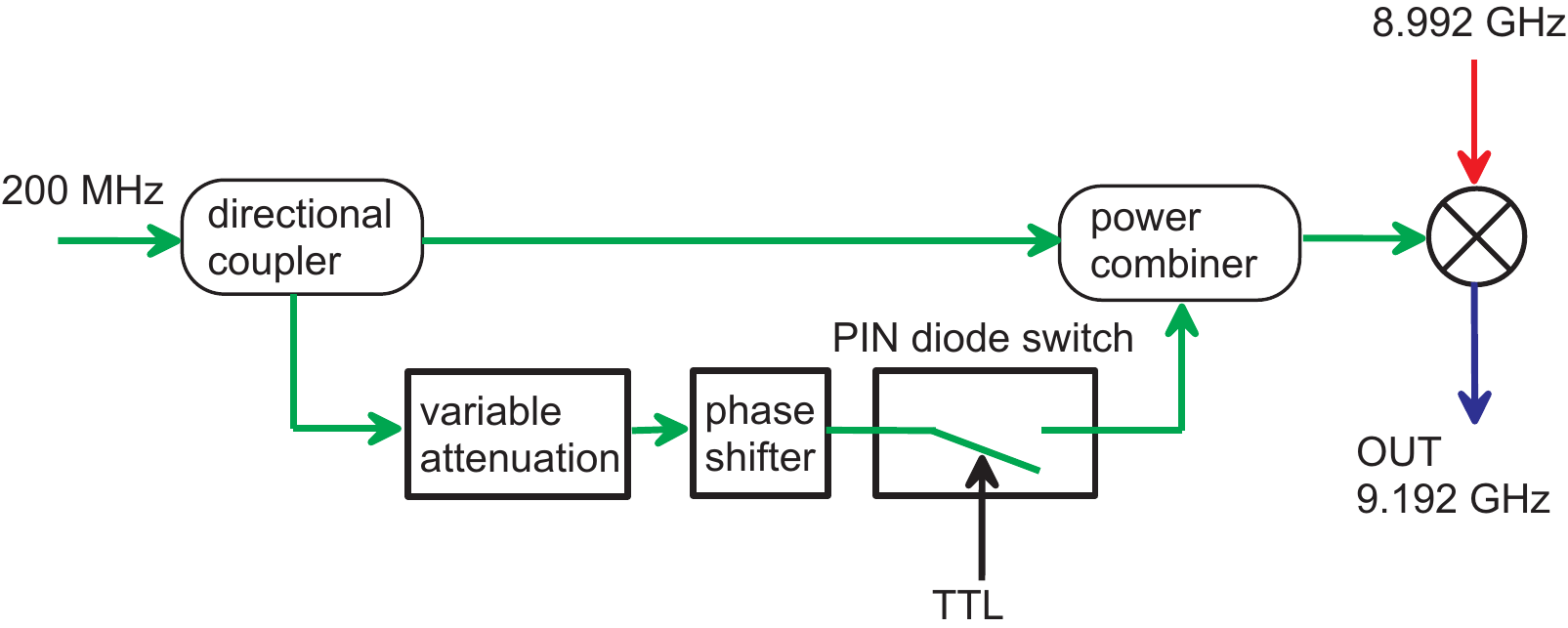}
\caption{Schematic of a switchable microwave synthesizer with a low phase transient using a Mach-Zehnder interferometric
switch.\label{fig_synthesizer}}
\end{figure}

Our latest experimental approach to evaluate microwave leakage shifts uses microwave synthesizers, which not only will have intrinsically low microwave leakage but also
can be switched off when the atoms are outside the microwave cavity. The challenge is to develop a switch that does not cause a large phase transient of the transmitted field after it is switched on. Figure \ref{fig_synthesizer} depicts the  architecture of our synthesizers. The interrogation signal at $9.192$~GHz is generated by mixing a high power $8.992$~GHz with a small amount of $200$~MHz signal to produce only the minimum needed power.

This architecture provides an ability to switch the carrier at the output by switching the $200$~MHz signal. To minimize phase transients, the $200$~MHz is switched using a Mach-Zehnder interferometric switch. The attenuator and the phase shifter are adjusted so that, when the PIN diode switch is closed, letting the signal pass through, destructive interference gives a dark fringe fringe at the output of the interferometer. When the PIN diode switch is open, the on-state signal passes the direct channel of the interferometer with no component that is likely to produce a phase transient. Specifically, the phase transient occurring in the PIN diode as it is switched is circumvented.

The development of a heterodyne phase transient analyzer \cite{Santarelli2009} was crucial to the successful implementation of switchable microwave synthesizers in Fig. \ref{fig_synthesizer} with low phase transients.
This analyzer was used extensively to select components and identify the critical parameters
to achieve the desired performance. Furthermore, this device can also be used to unveil spurious phase perturbations that are synchronous with
the fountain cycle, with sensitivities as low as $1~\mu$rad.s$^{-1}$ at the $9.192$~GHz carrier frequency.
Such perturbations can occur if some elements of the
fountain sequence (switching of fairly high levels of RF power in acousto-optic modulators, switching of the state selection microwave synthesizer, etc) perturb
the interrogation synthesizer through ground loops or other electromagnetic compatibility weaknesses. Required for the heterodyne phase transient analyzer, an independent reference synthesizer, the $6-9$~GHz synthesis ``for other purposes" in Fig. \ref{fig_ClockEnsemble}, is
distributed through an actively compensated fiber link.

With a suitably characterized switchable synthesizer, it becomes possible to test for a putative microwave leakage by measuring the clock frequency
with and without switched microwaves \cite{Santarelli2009}. All the LNE-SYRTE fountains (including FO2-Rb) are now equipped with a switchable synthesizer. Furthermore, synchronous perturbations are periodically searched for with the heterodyne phase transient analyzer.

\subsection{Accuracy budget}

Our approach to deal with other systematic shifts have been reported elsewhere and have not changed significantly since our last reviews \cite{Bize2004}\cite{Bize2005}.

In FO1 and FO2-Cs, the large cold collision shift \cite{Gibble1993} is extrapolated to zero density in real time through interleaved measurements at a high density and a halved density. We use the interrupted adiabatic passage method \cite{Pereira2002} which gives accurate ratios of the atomic densities. Owing to the much lower atom number dependent effects in $^{87}$Rb \cite{Fertig2000}\cite{Sortais2000}\cite{Bize2001}, we change the atom number and density with the more conventional ways, either loading time or Rabi flopping in the state selection cavity.
It is also possible to adopt an approach where a constant atom number is used most of the time, the collision shift being corrected based on a previously measured coefficient. The coefficient is measured with dedicated comparisons of high and low density, performed a few times per year. This approach is also used for FOM, which has a comparatively low cold collision shift due to a smaller atom number. Note that recently, another method has been proposed and demonstrated in order to deal with the cold collision shift of Cs \cite{Szymaniec2007}\cite{Szymaniec2011}.

The accuracy budget for the LNE-SYRTE fountains is given in Table \ref{tab_accuracy}. We give the most recent status, taking into account the latest evaluations of the distributed cavity phase shift and microwave lensing frequency shifts as reported in Table \ref{tab_DCP} and \ref{tab_MicrowaveLensing_a}.
In the absence of an independent calculation and/or experimental investigation, the uncertainty for the microwave lensing is conservatively taken to be equal to the calculated value. The last row of Table \ref{tab_accuracy} gives the status in accuracy budgets prior to 2011.

\begin{table*}
\renewcommand{\arraystretch}{1.3}

\begin{center}
\begin{tabular}{|l|c|c|c|c|}
\cline{2-5}
\multicolumn{1}{l|}{}                           &{\small  FO1}      & {\small FO2-Cs}   & {\small FOM}      & {\small FO2-Rb}   \\
\cline{2-5} \hline
{\small Quadratic Zeeman shift}                & $-1274.5 \pm 0.4$         & $-1915.9\pm0.3$           & $-305.6\pm 1.2$    & $-3465.5\pm0.7$  \\
\hline
{\small Blackbody radiation}                    & $172.6 \pm 0.6$           & $168.0\pm0.6$             & $165.6\pm0.6$      &   $122.8\pm1.3$\\
\hline
{\small Collisions and cavity pulling}     & $70.5\pm 1.4$                  & $112.0\pm 1.2$            & $28.6 \pm 5.0$     &   $2.0\pm2.5$\\
\hline
{\small Distributed Cavity Phase shift}  & $-1.0\pm 2.7$          &$-0.9\pm 0.9 $           &$-0.7\pm 1.6$               & $0.4\pm 1.0$   \\

\hline
{\small Spectral purity \& leakage}             & $<1.0$                    & $<0.5$                    &$<4.0$                   &   $<0.5$ \\
\hline
{\small Ramsey \& Rabi pulling}                 &$<1.0$                     &$<0.1$                     &$<0.1$               & $<0.1$  \\
\hline
{\small Microwave lensing}                      &$-0.7\pm0.7$                     &$-0.7\pm0.7$                     &$-0.9\pm0.9$             &  $-0.7\pm0.7$   \\

\hline
{\small Second order Doppler shift}            &$<0.1$                     &$<0.1$                     &$<0.1$            &   $<0.1$ \\
\hline
{\small Background collisions}                  & $<0.3$                    & $<1.0$                    & $<1.0$              &   $<1.0$ \\

 \hline \hline
{\small Total}                                  & $-1033.1\pm 3.5$           & $-1637.5\pm 2.1$          & $-113.0\pm 6.9$            &  $-3341.0\pm 3.3$  \\
                          \hline \hline

{\small Prior to 2011 (see note \cite{CommentAccuracy})}  & $-1031.4\pm4.1$  & $-1635.9\pm 3.8$ & $-111.4\pm8.1$   &$-3340.7\pm4.2$                   \\
\hline
\end{tabular}
\end{center}
\caption{Systematic fractional frequency corrections for FO1, FO2 and FOM, in units of $10^{-16}$. The table gives the most recent values taking into account the results of Tables \ref{tab_DCP} and \ref{tab_MicrowaveLensing_a}. The last row gives the overall corrections and uncertainties prior to 2011 (details in note \cite{CommentAccuracy}).
}
\label{tab_accuracy}
\end{table*}

\section{Applications in time and frequency metrology}

\subsection{$^{133}$Cs Fountain comparisons}

For more than a decade, the development of LNE-SYRTE fountain ensemble of Fig. \ref{fig_ClockEnsemble} has allowed high-accuracy comparisons between atomic fountain clocks. This has especially been the case in the last few years during which hundreds of days of clock comparisons have accumulated.

To deal with
the differences in clock cycles, mode of operation, schedule for verifying the magnetic field, etc, we apply the following data processing scheme.
Each fountain produces measurements, corrected for all systematic biases, at a rate of 1 measurement per cycle.
When an interleaved sequence of high and low density measurements is used, this first step requires to use these interleaved data to determine
the cold collision shift coefficient for a particular time window, and then to correct the high and low density frequency measurements accordingly.

The individual fountain measurements are averaged over time intervals of 864~s (1/100th of a day) starting at the beginning of the day.
Intervals are discarded if the number of valid fountain cycles is below a chosen threshold, for instance, due to an automated verification of the magnetic field, a laser going out of lock, or a phase/frequency jump in the output of the cryogenic oscillator. Other filters can be applied based on fault of the CSO, one of the fiber links, or other subsystems. This leaves a set of validated data where the sampling sequence is matched between all fountains. These data represent measurements by each fountain of the frequency of the common ultra-stable reference, i.e. the 11.98~GHz weakly phase locked to one of the hydrogen masers (see Fig. \ref{fig_ClockEnsemble}), which are used for calibration of the international atomic time (see \ref{subsec_TAI} below).

\begin{figure}[h]
\includegraphics [width= \linewidth]{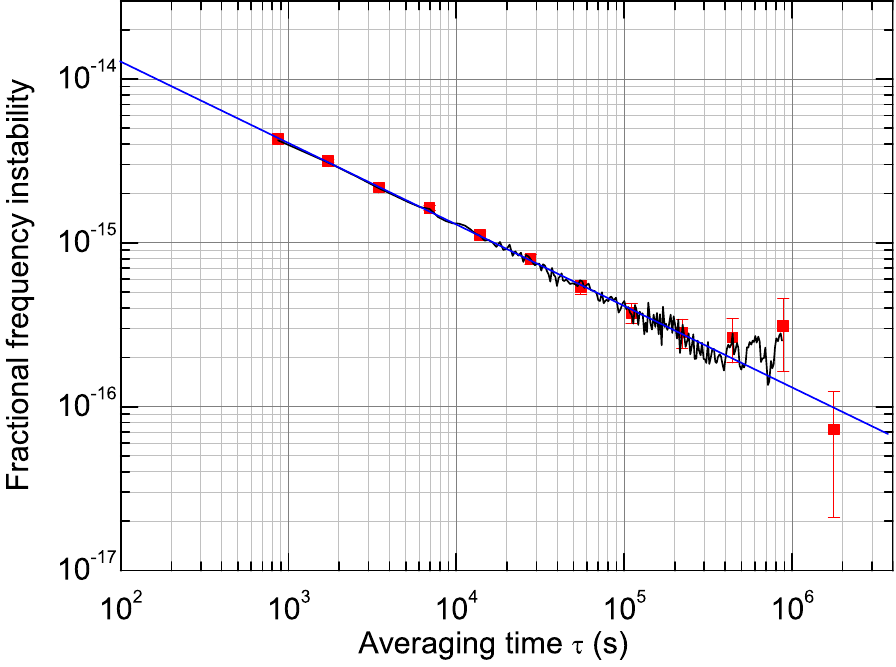}
\caption{Fractional frequency instability between the FO2 and FOM Cs fountains as a function of the averaging time $\tau$. The data corresponds to 50 effective days of comparisons between MJD 55551 and MJD 55663. Red points with error bars gives the instability at $\tau=2^{N}\times 864$~s. The black curve shows the instability for all $\tau$.\label{fig_StabilityFOMFO2}}
\end{figure}

As a third step, the frequency difference between the synchronous validated data of two fountains eliminates the frequency of the common-mode ultra-stable reference and allows a direct comparison. An example of frequency instability between FOM and FO2 over 50 effective days of averaging is shown in Fig. \ref{fig_StabilityFOMFO2}. Here, the overall short-term instability is $1.3\times 10^{-13}\tau^{-1/2}$ and the measurement resolution is $\sim 2$ part in $10^{16}$ after averaging for $4.5$~days. The best reported instability for this type of comparison is $5.0\times 10^{-14}\tau^{-1/2}$ \cite{Bize2004}\cite{Bize2005}\cite{Vian2005}.

\begin{table*}
\renewcommand{\arraystretch}{1.3}

\begin{center}
\begin{tabular}{|l|c|c|c|c|c|c|c|c|}
\cline{2-9}
\multicolumn{1}{l|}{}     &    {\small  Duration (days)}                  &{\small  $D$}      & {\small $U$}   & {\small $|D/U|$}      & {\small $\chi_D^2$} & {\small $Q_D$} & {\small $\chi_0^2$} & {\small $Q_0$}\\

\cline{2-9}
\hline
{\small FOM-FO2}       & $320$         & $-3.0$         & $9.1$           & $0.33$    & $0.28$  & $0.996$ & $0.37$ & $0.983$ \\
\hline
{\small FOM-FO1}   & $240$  & $4.1$                  & $9.2$            & $0.44$  & $0.12$   &   $0.999$ & $0.28$ & $0.989$ \\
\hline
{\small FO1-FO2}  & $500$ & $-5.7$          &$6.2$           &$0.92$               & $1.48$ & $0.55$ & $2.12$ & $0.001$\\

\hline
\end{tabular}
\end{center}
\caption{Summary of Cs fountain comparisons over the November 2006 to April 2011 period. The first column gives the cumulated number of days of comparison. $D$ is the fractional frequency difference and $U$ the overall uncertainty, both in units of $10^{-16}$. $\chi_D^2$ is the reduced $\chi^2$ with respect to the mean difference $D$ . $\chi_0^2$ is the reduced $\chi^2$ with respect to the expected null difference. $Q_D$ and $Q_0$ are the corresponding goodness-of-fit as defined in \cite{NR1992}.
}
\label{tab_comparison}
\end{table*}

\begin{figure}[h]
\includegraphics [width= \linewidth]{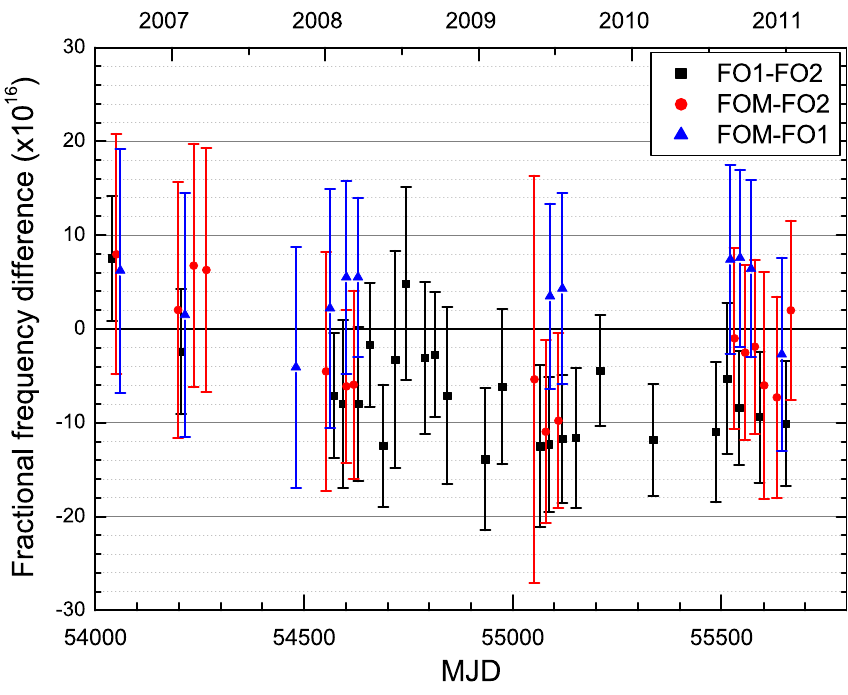}
\caption{Fractional frequency differences between the three pairs of LNE-SYRTE Cs fountains from November 2006 to April 2011. Error bars are the total uncertainties which are dominated by systematic uncertainties.}\label{fig_DiffFontaines}
\end{figure}

Fountain comparisons following the above data processing scheme are now performed regularly at LNE-SYRTE.
It was recently incorporated into a fully automated procedure. In addition to daily updates of the fountain comparisons, an overview of the systems' parameters is generated to help assess the status of the clock ensemble in quasi-realtime with hourly updates. To produce fully validated data for fountain comparisons, for example for timescale steering and TAI calibrations, the entire data processing is critically scrutinized and redone if necessary. Figure \ref{fig_DiffFontaines} shows the monthly average of fractional frequency differences $D$ between the three Cs fountain pairs from November 2006 to April 2011. These data correspond to a total of $\sim 1000$ effective days of fountain comparison. The error bars are the total uncertainties for a given clock pair, which are dominated by the systematic uncertainties of the two fountains. The overall fractional frequency differences $D$ (weighted averages) are reported in Table \ref{tab_comparison}, together with the corresponding overall uncertainties $U$. $|D/U|$ represents the deviation from a null difference scaled to the $1\sigma$ total uncertainty.
The reduced $\chi^{2}$, either with respect to the mean ($\chi_D^2$) or with respect to zero ($\chi_0^2$) gives a further consistency check of these data.
Finally, the goodness-of-fit\footnote{The goodness-of-fit $Q$ is the probability that a value of chi-square as poor as the value found with the data should occur by chance. Quoting \cite{NR1992}: ``If $Q$ is larger than, say, 0.1, then the goodness-of-fit is believable".} as defined in \cite{NR1992} can be computed, either with respect to the mean ($Q_D$) or with respect to zero ($Q_0$). The $Q$'s very close to 1 for FOM-FO2 and FOM-FO1 indicate that fluctuations about the means are smaller than the total uncertainties. $Q_0$ which tests deviations (and/or fluctuations) from the expected null difference shows that FOM-FO2 and FOM-FO1 agree within the total uncertainties. Instead the low $Q_D$ and $Q_0$ for the more stringent FO1-FO2 comparison would indicate a slightly underestimated systematic effect, of a few $10^{-16}$, which is under investigation.
Note that for the data of Fig. \ref{fig_DiffFontaines}, the modifications from our recent study of the DCP shift (see \ref{subsec_DCP} and \cite{Guena2011}) were not yet implemented.
Consequently, the DCP shift is one potential candidate to explain these deviations. Future comparisons, after a full implementation to all fountains of the modified approach to the DCP shift, will clarify this point.

Note that a somewhat similar analysis of worldwide comparisons of fountain primary frequency standards exploiting their calibrations of TAI is reported in \cite{Parker2010}.

\subsection{A secondary representation of the SI second}

Following the early demonstration of the low cold collision shift in $^{87}$Rb \cite{Fertig2000}\cite{Sortais2000}, LNE-SYRTE has performed a series of high accuracy absolute measurements of the $^{87}$Rb hyperfine frequency against Cs fountain clocks \cite{Bize1999}\cite{Marion2003}\cite{Guena2010}. The 2003 comparison was the most accurate of any atomic frequency comparison at the time. Since then, it has been superseded by the rapidly improving optical frequency measurements. The $^{87}$Rb hyperfine frequency was the first secondary representation of the SI second recognized by BIPM \cite{CCTF2004}. Its recommended value is given by the 2002 LNE-SYRTE measurement: 6~834~682~610.904324~Hz with a recommended uncertainty of $21~\mu$Hz or $3$ parts in $10^{15}$.
This recommended uncertainty was chosen to be $\sim 3$ times larger than the actual uncertainty of the measurement ($1.3\times 10^{-15}$). Our latest reported measurement ($4~\mu$Hz uncertainty or $5.9$ parts in $10^{16}$), together with an analysis of the consistency of all our measurements can be found in \cite{Guena2010}.
An independent $^{87}$Rb \emph{versus} $^{133}$Cs fountain comparison has been reported for the first time in 2010. Agreement with LNE-SYRTE measurements is claimed at the $10^{-15}$ level but
the actual result of the measurement has not yet been reported \cite{Ovchinnikov2011}\cite{Li2011b}\cite{Ovchinnikov2011a}. Also notable is the construction of at least 8 Rb fountains at the US Naval Observatory \cite{Peil2006}\cite{Peil2011}.

\subsection{Contributions to TAI}\label{subsec_TAI}

\begin{figure}[h]
\includegraphics [width= \linewidth]{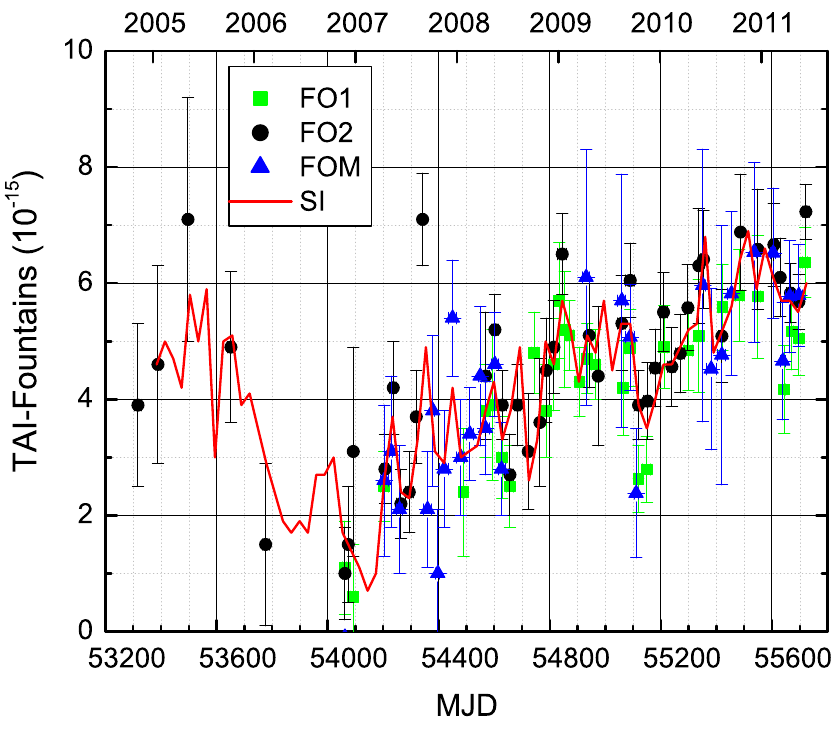}
\caption{Calibrations of the International Atomic Time (TAI) by the LNE-SYRTE fountains since 2004. The data are from BIPM \emph{circular T}. Each point is the
fractional frequency difference between TAI and the SI second as measured by one of LNE-SYRTE's fountains and corresponds to a formal monthly report to BIPM. The red curve is the difference between TAI and the SI second, as computed by BIPM from all available primary standards.
}\label{fig_TAI}
\end{figure}

Since 2007, LNE-SYRTE fountains have measured the frequency of one of LNE-SYRTE H-masers almost continuously.
These data are used to calibrate the International Atomic Time (TAI).
Figure \ref{fig_TAI} shows all calibrations of TAI by LNE-SYRTE cesium fountains from November 2004. Each data point corresponds to a measurement, by one fountain, of the H-maser average frequency over a $20-30$ days period, reaching a typical statistical uncertainty of $3\times 10^{-16}$,
which is augmented in quadrature by an uncertainty due to dead times of typically $1.5\times 10^{-16}$ \cite{CircularT}. Between the 2007 and 2010 period, LNE-SYRTE reported 78 formal
evaluations to BIPM, out of a total of 161 fountain reports worldwide, i.e. LNE-SYRTE has contributed almost $50\%$ of
all reports since 2007. For contributions to TAI, the gravitational redshift corrections are $-69.3\times 10^{-16}$, $-65.4\times 10^{-16}$ and $-68.7\times 10^{-16}$ for FO1, FO2 and FOM respectively, with an uncertainty of $1.0\times 10^{-16}$ corresponding to a $1$~m uncertainty of height above the geoid.
Contributions to TAI also provide a vehicle to compare to other fountain clocks around the world. The latest update on these comparisons appears in \cite{Parker2010}.

Since mid-2006, fountain data, including from FO2-Rb, are used to steer the French Atomic Time TA(F) on a monthly basis. TA(F) is a timescale based on an ensemble of commercial Cs clocks located in France and elaborated at LNE-SYRTE.
This steering has improved the long-term stability of the timescale with respect to TAI \cite{Uhrich2008}.

\subsection{Absolute measurements of optical frequencies}

The accuracy and stability of optical atomic clocks have now largely surpassed those of microwave clocks, and notably atomic fountain clocks \cite{Chou2010}\cite{Lemke2009a}\cite{Rosenband2008}\cite{Ludlow2008}. This opens a path to redefining of the SI second with an optical transition \cite{CGPM2009}. One of the prerequisites to a redefinition is absolute frequency measurements of potential optical transitions with the highest accuracy. The LNE-SYRTE fountain clock ensemble (Fig. \ref{fig_ClockEnsemble}) offers some unique features to contribute to this task. This includes the availability of several fountains for redundant measurements, of an ultra-low noise reference signal that provides the best reported stability to date between a microwave and optical clock \cite{Baillard2007b}, and of the transportable fountain FOM for measuring optical frequency at other laboratories. Absolute frequency measurements of optical frequencies can also probe the stability of fundamental constants \cite{Blatt2008}\cite{Fischer2004}\cite{Fortier2007}\cite{Peik2004}, determine the Rydberg constant \cite{Mohr2008} and test Quantum Electrodynamics (QED) \cite{Karshenboim2005}.

Next, we give an overview of absolute optical frequency measurements made with LNE-SYRTE fountain ensemble.


\subsubsection{Hydrogen $\mathrm{1S-2S}$ transition measured with $\mathrm{FOM}$ at $\mathrm{MPQ}$ Garching, Germany}\label{subsec_H1S2S}

At the Max Planck Institut f\"ur Quantenoptik (MPQ) in Garching (Germany), we have measured the absolute frequency of the 1S-2S optical transition via 2 photon excitation at 246~nm \cite{Niering2000}\cite{Fischer2004}\cite{Parthey2011a}. Three measurement campaigns were completed in 1999, 2003 and 2010.
For these experiments a femto-comb laser is used to measure the frequency of the excitation laser. The repetition rate is locked to a hydrogen maser whose frequency is simultaneously measured directly by FOM. The noise of the maser signal is filtered by phase-locking a BVA quartz oscillator to improve the measurement stability. The last measurement in 2010 had the best fractional uncertainty: $4.2\times 10^{-15}$ \cite{Parthey2011a}. The measured 1S-2S transition frequency is 2~466~061~413~187~035(10) Hz.

 High resolution spectroscopy of hydrogen is interesting to test highly accurate atomic structure calculations for QED tests \cite{Pohl2010}. Also, the recent capture of antihydrogen at CERN \cite{AlphaCollaboration2011} is a first step toward cooling of antimatter to enable high resolution spectroscopy of
antihydrogen for fundamental tests of charge, parity, time reversal (CPT) symmetry and the gravity of antimatter.

\subsubsection{$^{40}\mathrm{Ca}^{+}$ optical clock measured with $\mathrm{FOM}$ at the University of Innsbruck, Austria}

In 2007 at the Institut f\"ur Experimentalphysik and Quantenoptik, University of Innsbruck (Austria), we measured the absolute frequency of the $4s~^{2}S_{1/2}-3d~^{2}D_{5/2}$ optical quadrupole transition of a trapped single $^{40}$Ca$^{+}$ ion. The frequency of this transition is 411~042~129~776~393.2 (1.0)~Hz \cite{Chwalla2009}. This corresponds to a fractional frequency uncertainty of 2.4 parts in $10^{15}$. During this experiment, FOM was used in the autonomous configuration mentioned in \ref{subsec_FOM}, where a BVA quartz oscillator is frequency locked to the FOM spectroscopic signal. The output of the BVA quartz oscillator synchronizes the repetition rate of the femto-comb laser that measures the probe laser frequency for the $^{40}$Ca$^{+}$ transition. This quadrupole transition in $^{40}$Ca$^{+}$ ion is another candidate for an optical clock and for tests of the stability of fundamental constants.

\subsubsection{$^{87}\mathrm{Sr}$ and $^{88}\mathrm{Sr}$ optical lattice clocks measured at $\mathrm{SYRTE}$}\label{subsec_Sr}

LNE-SYRTE is developing two Sr optical lattice clocks. During their development, several absolute frequency measurements of the Sr clock frequency were made \cite{Baillard2007b}\cite{Baillard2007}\cite{LeTargat2006}\cite{Courtillot2003}. The best reported uncertainty for these measurements is $2.6\times 10^{-15}$ \cite{Baillard2007b}. Both Titanium:Sapphire based and Er-doped fiber based optical frequency combs were used to measure the frequency of the 698~nm laser light stabilized to the Sr atoms. The 2006 measurements of the $^{87}$Sr used a fiber optical frequency comb transported from the Physikalisch Technische Bundesanstalt (PTB) in Braunschweig (Germany) \cite{LeTargat2006}\cite{Baillard2007b}.
As already mentioned, the ultra-stable reference based on a CSO (Fig. \ref{fig_ClockEnsemble}) allows the best reported short-term stability to date for comparing an optical clock to a microwave clock: $6\times 10^{-14}\tau^{-1/2}$. In the recent measurements, the optical frequency comb is stabilized to an ultra-stable laser as described in \cite{Dawkins2010}, making the repetition rate of the femtosecond laser an ultra-stable microwave signal. This signal is then measured against the CSO based ultra-stable reference which is in turn measured by the fountain clocks. Simultaneously, the beat between the optical frequency comb and the 698~nm light is counted to extract its absolute frequency. Combining these with similar measurements by other institutes, stringently tests the stability of fundamental constants \cite{Blatt2008}.

\subsubsection{Measurement of $^{199}\mathrm{Hg}$ and $^{201}\mathrm{Hg}$ optical clock transition frequencies at $\mathrm{SYRTE}$}

LNE-SYRTE is developing an optical lattice clock based on neutral mercury. The first absolute frequency measurements of the $^{199}$Hg and $^{201}$Hg optical clock transition were in 2008 \cite{Petersen2008a}. These initial measurements improved the previous knowledge of the transition frequency by more than 4 orders of magnitude, to a fractional frequency uncertainty of $5\times 10^{-12}$. More recently, measurements against LNE-SYRTE ultra-stable reference were used to perform the first experimental determination of the magic wavelength for Hg \cite{Yi2011}. Ongoing development of the Hg clock is expected to soon yield measurements at the $10^{-15}$ level and beyond.

\subsection{Other measurements with the transportable fountain $\mathrm{FOM}$}

We have performed frequency comparisons with the engineering model of the space clock PHARAO (see \ref{subsec_PHARAO} below and \cite{Laurent2006}) between 2007 to 2009, at the Centre National d'Etudes Spatiales (CNES) in Toulouse, France.
We have also verified the frequency performance of the ACES architecture \cite{Cacciapuoti2007}. The test included a ground model of the onboard data handling unit (XPLC), the Frequency Comparison and Distribution Package (FCDP), the PHARAO engineering model, and a ground model of the Space Hydrogen Maser (SHM). The two clocks are combined to generate a timescale with the short-term stability of SHM and the long-term stability and accuracy of PHARAO.

\subsection{Remote comparisons via T2L2 satellite laser link}
In 2010, time transfer by the T2L2 laser link \emph{via} the JASON2 satellite \cite{Guillemot2008} was tested by comparing FOM at the Observatoire de la C\^ote d'Azur (OCA) in Grasse (France) with the other LNE-SYRTE fountains at the Observatoire de Paris. The common time transfer techniques, carrier phase GPS and TWSTFT, were also used in parallel. Notable in this comparison is the large gravitational redshift, $-1.384\times 10^{-13}$, at the OCA altitude of 1268~m.
Preliminary results \cite{Samain2011} show that all the time transfer methods are consistent to within 2~ns over 2 months. The frequency difference between the remote fountains is measured with a typical uncertainty of $1\times 10^{-15}$, which determines the gravitational redshift difference to $10^{-2}$.
It now remains to refine the analysis and to evaluate the ultimate performances of the T2L2 time transfer link.
Note that at MPQ, at CNES and at OCA, FOM also provided calibration of the TAI by using a GPS receiver and a carrier phase GPS analysis software developed by CNES or by NRCAN. This is the first time that a primary frequency standard contributes to steering the TAI from different sites with different gravitational redshift \cite{Rovera2011}.

\section{Applications in fundamental physics}

One of the most interesting applications of atomic clocks is to test fundamental physics. In this section, we review contributions of LNE-SYRTE fountain ensemble to such tests.

\subsection{Stability of constants}

Repeated high accuracy comparisons between atomic (or molecular) frequencies can test the stability of fundamental constants such as the fine structure constant $\alpha$ or the electron to proton mass ratio $m_e/m_p$ (see, for instance, \cite{Karshenboim2000b} \cite{Flambaum2008a} and references therein). Laboratory experiments usefully complement tests over cosmological timescales since the interpretation does not rely on any assumptions about a cosmological model \cite{Uzan2003}\cite{Uzan2011}.

A first test performed with LNE-SYRTE fountain ensemble comes from a series of $^{87}$Rb \emph{versus} $^{133}$Cs hyperfine frequency comparisons made during the development of the FO2 dual Rb/Cs fountain \cite{Marion2003}\cite{Bize2004}\cite{Bize2005} and \cite{Guena2008}. In this last 2008 report, a putative time variation of the ratio of the two hyperfine frequencies is constrained to $d\ln(\nu_{\mathrm{Rb}}/\nu_{\mathrm{Cs}})/dt=(-3.2\pm 2.3)\times 10^{-16}$~yr$^{-1}$. In terms of fundamental constants, this result yields $d\ln(\alpha^{-0.49} [g_{\mathrm{Rb}}/g_{\mathrm{Cs}}])/dt=(-3.2\pm 2.3)\times 10^{-16}$~yr$^{-1}$ where $g_{\mathrm{Rb}}$ and $g_{\mathrm{Cs}}$ are the nuclear g-factors in $^{87}$Rb and $^{133}$Cs. Expressing these g-factors in terms of fundamental parameters of the Standard Model \cite{Flambaum2004}\cite{Flambaum2006} gives: $d\ln(\alpha^{-0.49} [m_q/\Lambda_{\mathrm{QCD}}]^{-0.025})/dt=(-3.2\pm 2.3)\times 10^{-16}$~yr$^{-1}$, where $\Lambda_{\mathrm{QCD}}$ is the mass scale of Quantum Chromodynamics (QCD). An improved analysis including more recent measurements, with FO2 in the dual fountain configuration, will improve this value significantly.

A second test comes from a series of absolute frequency measurements of the $^{87}$Sr optical lattice clock by LNE-SYRTE (see \ref{subsec_Sr}), the University of Tokyo, Japan and JILA, Boulder, Colorado, USA \cite{Blatt2008}. Together, these constrain the putative variation of $\nu_{\mathrm{Sr}}/\nu_{\mathrm{Cs}}$ to $d\ln(\nu_{\mathrm{\mathrm{Sr}}}/\nu_{\mathrm{\mathrm{Cs}}})/dt=(-7\pm 18)\times 10^{-16}$~yr$^{-1}$ corresponding to $d\ln(\alpha^{2.77}[m_q/\Lambda_{\mathrm{QCD}}]^{-0.039}[m_e/\Lambda_{\mathrm{QCD}}])/dt=(7\pm 18)\times 10^{-16}$~yr$^{-1}$. The same measurements constrain the putative variation of this same combination of constants with the gravitational potential to $c^2 d\ln(\alpha^{2.77}[m_q/\Lambda_{\mathrm{QCD}}]^{-0.039}[m_e/\Lambda_{\mathrm{QCD}}])/dU=(-5.8\pm 8.9)\times 10^{-6}$, where $c$ is the speed of light and $U$ the gravitational potential \footnote{For instance, the gravitational potential created by a point mass $m$ at a distance $r$ is: $U(r)=G m/r$ where $G$ is the Newtonian constant of gravitation.}. This test relies on exploiting the yearly modulation of the gravitational potential of the Sun due the eccentricity of the Earth orbit.

Measurements of the H(1S-2S) transition performed at MPQ Garching (see \ref{subsec_H1S2S}) using the transportable fountain FOM as a reference, offer a third test: $d\ln(\nu_{\mathrm{H(1S-2S)}}/\nu_{\mathrm{Cs}})/dt=(-3.2\pm 6.3)\times 10^{-15}$~yr$^{-1}$ \cite{Fischer2004}. This translates to $d\ln(\alpha^{2.83}[m_q/\Lambda_{\mathrm{QCD}}]^{-0.039}[m_e/\Lambda_{\mathrm{QCD}}])/dt=(3.2\pm 6.3)\times 10^{-15}$~yr$^{-1}$, a constraint which will soon improve as a result of a recent measurement campaign \cite{Parthey2011a}.

A fourth test comes from a series of comparisons between the $^{133}$Cs and the hydrogen hyperfine frequencies. LNE-SYRTE fountain's contributions to this test were the calibrations of TAI reported in BIPM \emph{Circular T} (see \ref{subsec_TAI}) enabling the connection with 4 hydrogen masers participating to the elaboration of the NIST AT1 atomic timescale (see for instance \cite{Parker1997}). Other Cs fountain contributions to the experiment came from PTB in Germany and from the Istituto Nazionale di Ricerca Metrologica (INRIM) in Italy. The analysis was performed at the National Institute of Standards and Technology (NIST) Boulder, Colorado, USA \cite{Ashby2007} and gives $|c^2 d\ln(\nu_{\mathrm{H}}/\nu_{\mathrm{Cs}})/dU|=(0.1\pm 1.4)\times 10^{-6}$, corresponding to $|c^2 d\ln(\alpha^{0.83}[m_q/\Lambda_{\mathrm{QCD}}]^{0.11})/dU|=(0.1\pm 1.4)\times 10^{-6}$.

Note that instead of using $\alpha$, $m_q/\Lambda_{\mathrm{QCD}}$ and $m_e/\Lambda_{\mathrm{QCD}}$ as independent variables, $\alpha$, $\mu=m_e/m_p$ and $m_q/m_p$ can be used. The link between the two approaches is $d\ln(m_p/\Lambda_{\mathrm{QCD}})\simeq 0.048\times d\ln(m_q/\Lambda_{\mathrm{QCD}})$ \cite{Flambaum2006}.

\subsection{Test of Lorentz Invariance in the matter sector}

The LNE-SYRTE FO2 fountain has tested the anisotropy of space as in Hughes-Drever experiments (see for instance \cite{Will2006}). For this experiment, the fountain sequence is tailored to probe opposing Zeeman transitions in $^{133}$Cs to test for a putative variation of their frequencies when the orientation of the quantization axis changes (here, due the Earth rotation) with respect to a supposedly preferred frame \cite{Wolf2006}, such as the frame of the Cosmic Microwave Background. The experiment was interpreted within the framework of the Lorentz violating Standard Model Extension (SME), where it is sensitive to proton parameters corresponding to a largely unexplored region of the SME parameter space.
The constraints for 4 parameters, already constrained by other measurements, were improved by as much as 13 orders of magnitude and 4 parameters were constrained for the first time. Operating FO2 in the dual Rb/Cs configuration \cite{Guena2010} opens new possibilities for improving and complementing these tests.

\subsection{Other tests}

The development of LNE-SYRTE fountain ensemble also offered a test of Lorentz Local Invariance in the photon sector, not using the fountains, but simply comparing the CSO to a hydrogen maser (see Fig. \ref{fig_ClockEnsemble}) for a duration that now exceeds 10 years. This experiment is to date the most stringent Kennedy-Thorndike test by a factor of $500$. The latest update is published in \cite{Tobar2010}. The experiment can also be interpreted in the SME framework \cite{Wolf2004a}.


\section{Development of the PHARAO cold atom space clock}\label{subsec_PHARAO}

\begin{figure}[h]
\includegraphics [width= \linewidth]{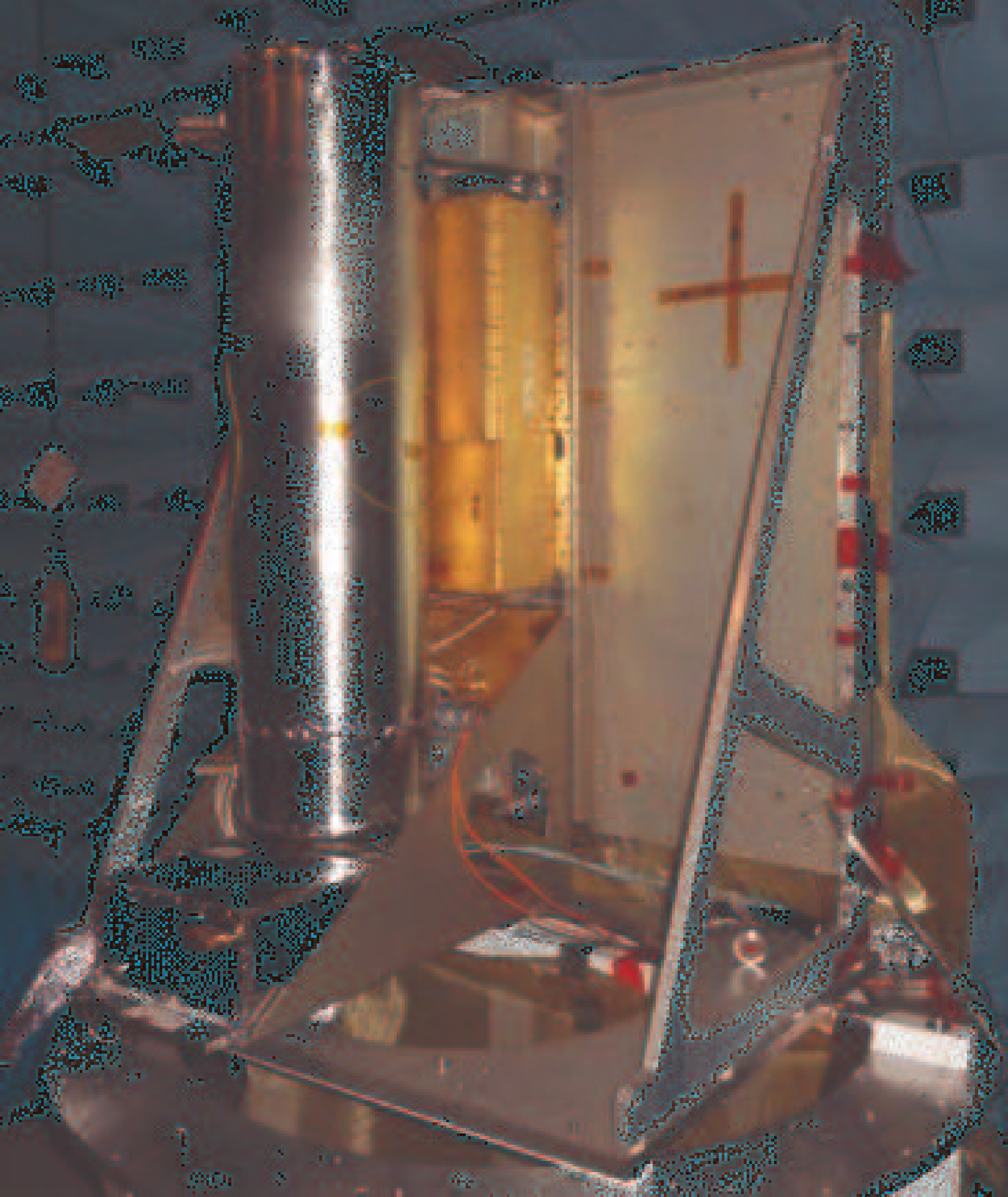}
\caption{The engineering model of the PHARAO cold atom space clock under test at CNES, Toulouse (photo is courtesy of CNES). The picture shows the vertically mounted ``cesium tube", the vacuum chamber surrounded by the magnetic shields. The gold-coated box behind the cesium tube contains the ``laser source". A second gold-coated box, below the laser source, is the microwave synthesizer. The black box in front, at the bottom of the cesium tube, is the on-board computer unit.}\label{fig_PHARAO}
\end{figure}

\begin{figure}[h]
\includegraphics [width= \linewidth]{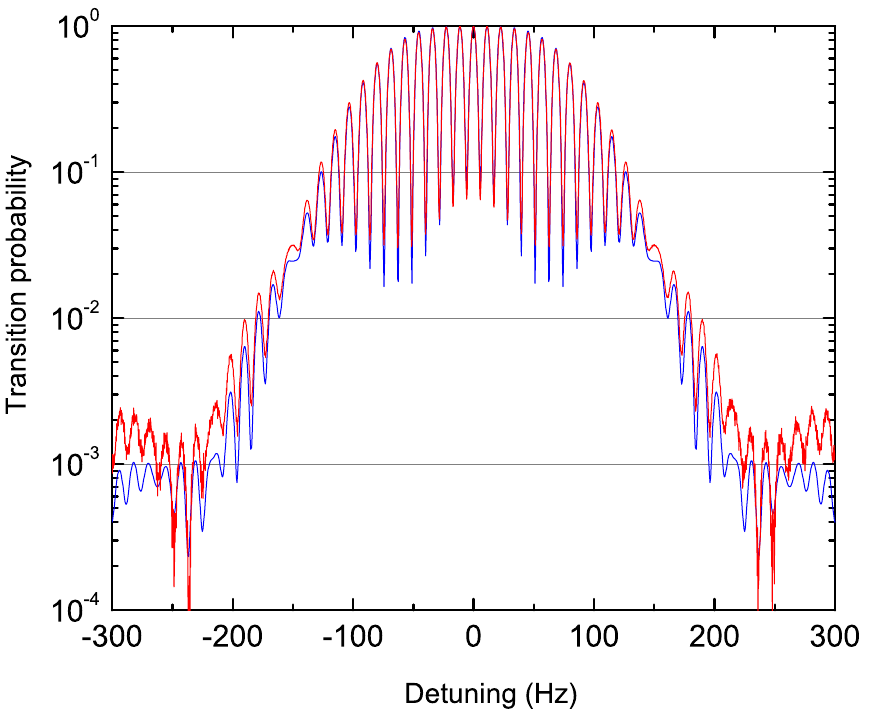}
\caption{Experimental (red) and calculated (blue) Ramsey fringes of the PHARAO-EM tested on Earth. The central fringe does not have the highest contrast because, operating on Earth, gravity leads to different interaction times on the two cavity interactions.\label{fig_fringesPHARAO}}
\end{figure}

Atomic Clocks Ensemble in Space (ACES) is a European Space Agency (ESA) fundamental physics mission originally proposed by Laboratoire Kastler Brossel and SYRTE.
It is based on the operation of highly stable and accurate atomic clocks in the microgravity environment of the International Space Station (ISS) \cite{Cacciapuoti2007}. The time scale generated by the ACES clocks on-board the ISS is delivered to Earth through a high-performance two-way time and frequency transfer link. The clock signal is used to perform space-to-ground as well as ground-to-ground comparisons of atomic frequency standards.
The ACES scientific objectives cover both fundamental physics and applications. Tests of Special and General Relativity will be performed with improved accuracy, as well as a search for temporal variations of fundamental constants. On the application side, frequency comparisons between distant clocks, both space-to-ground and ground-to-ground, will be performed worldwide with unprecedented resolution. ACES will demonstrate a new type of ``relativistic geodesy", which, based on a precision measurement of the Einstein's gravitational red-shift, will resolve differences in the Earth gravitational potential at the $10$~cm level. Finally, ACES will contribute to the improvement of global navigation satellite systems (GNSS) and to future evolutions of these systems. It will demonstrate new methods to monitor the ocean surface based on scatterometric measurements of the GNSS signal and it will contribute to the monitoring of the Earth atmosphere through radio-occultation experiments.
The expected performance is a time stability of 10~ps over ten days (1~ps for common view comparisons between ground clocks) and a frequency accuracy better than $3\times 10^{-16}$.
The accuracy and the long term frequency stability are defined by the PHARAO clock, an instrument developed by the French space agency CNES and SYRTE. The clock uses cold cesium atoms, slowly moving through a Ramsey cavity.
We have fully tested the engineering model (PHARAO-EM) of the clock \cite{Laurent2006}. Figure \ref{fig_PHARAO} shows a picture of the PHARAO-EM under test at the CNES assembly room in Toulouse.  For nominal operation the frequency stability is $3.3\times 10^{-13} \tau^{-1/2}$. The central Ramsey resonance linewidth is 5.6~Hz and $10^{6}$ atoms are detected.
The contrast is only $93\%$ (Fig. \ref{fig_fringesPHARAO}) since, on Earth, the atoms do not spend the same time in the two interaction zones of the Ramsey cavity.

On Earth the clock cycle duration lasts 790~ms with a microwave interrogation time of $90$~ms. Consequently the phase noise of the microwave signal, synthesized from a quartz oscillator, contributes to the frequency instability at a level of $2.2\times 10^{-13}\tau^{-1/2}$.
We have performed an accuracy evaluation of the clock. The main frequency corrections and their uncertainties are shown in table \ref{tab_accuracyPHARAO}. The total frequency uncertainty is $1.9\times 10^{-15}$.The largest uncertainty is the quadratic Zeeman shift because the magnetic field along the atomic trajectories is not homogeneous due to a problem with the innermost magnetic shield. The second largest contribution is the cold collision frequency shift. The frequency shift per detected atom is similar to other cold cesium clocks and the uncertainty is mainly limited by the measurement duration (only two weeks). The blackbody radiation shift is deduced from measurements of the ambient temperature. A technical problem prevented us from using the calibrated temperature probes located next to the Ramsey cavity. Finally, the first order Doppler frequency shift has a large value since the atom's axial velocity decreases due to gravity field, which causes the longitudinal DCP shifts from the 2 interactions to not cancel. This gives a large DCP shift that has a large dependance on the microwave field amplitude.

\begin{table}
\renewcommand{\arraystretch}{1.3}

\begin{center}
\begin{tabular}{|l|c|c|}
\cline{2-3}
\multicolumn{1}{l|}{}                           &{\small  Correction}       & {\small Uncertainty}      \\
\cline{2-3} \hline
{\small Quadratic Zeeman effect}                & $-5633$                  & $14$                     \\
\hline
{\small Blackbody radiation}                    & $162.5$                   & $2.2$                     \\
\hline
{\small Cold collisions \& cavity pulling}      & $63$                      & $11$                     \\
\hline
{\small First order Doppler}                    & $-33$                   & $5$                       \\

 \hline \hline
{\small Total}                                  & $-5440$                  & $19$                     \\
\hline
\end{tabular}
\end{center}
\caption{Systematic fractional frequency corrections for the PHARAO engineering model during the 2009 tests, in units of $10^{-16}$.}
\label{tab_accuracyPHARAO}
\end{table}

The PHARAO-EM clock, operated in several configurations, was compared to the transportable fountain FOM.
After applying all systematic corrections (Table \ref{tab_accuracy} and \ref{tab_accuracyPHARAO}), the frequency differences were smaller than $2\times 10^{-15}$, with a resolution of $5\times 10^{-16}$. During these measurements, the accuracies of the clocks were $8.1\times 10^{-16}$ for FOM and $1.9\times 10^{-15}$ for PHARAO-EM. Consequently, there is no significant frequency offset between the clocks. Testing the engineering model demonstrated the full operation of the PHARAO clock architecture and highlighted some improvements to implement in the flight model. The PHARAO flight model is now being assembled and it will be delivered in 2012 together with the other ACES instruments. The launch date will be 2 years after the delivery of all the instruments.

SYRTE will be a master station to survey and to analyze the ACES signal. One of the microwave link station developed for ACES will be installed on the roof of a building at the Observatoire de Paris. Hydrogen masers steered by the atomic fountains and UTC(OP) will synchronize the station, forming the ACES ground
segment at SYRTE. Strontium and mercury optical lattice clocks at SYRTE \cite{Westergaard2011}\cite{Yi2011}, which are connected to the fountain ensemble \emph{via} optical frequency combs, will also contribute to the ACES mission.
Notably, ACES will provide ground-to-ground comparisons of these optical clocks to similar clocks at other institutes around the world, demonstrating relativistic geodesy.

\section{Conclusions}

This paper gives an extensive overview of the development of the LNE-SYRTE fountain ensemble. We can also identify some priorities for the future. For fountains, it will be important to further improve inter-comparisons to reduce inaccuracies to 1 to 2 parts in $10^{16}$, and to possibly remove yet unidentified sources of uncertainty or instability. Implementing more sophisticated microwave cavity designs is one interesting approach. Simultaneously, the reliability of the fountain ensemble must further improve. This is required by the ground segment of the ACES mission - quasi-continuous operation at the best accuracy of the three fountains for at least 18 months. This will be achieved by a number of technical improvements in the fountains (notably the laser systems), the fountain environment, and by complementing the cryogenic sapphire oscillator with a second low-noise microwave source derived from an ultra-stable laser. This system will also ensure a permanent link to the optical clocks. Achieving these goals will enable new or improved applications: realization of an ultra-stable local timescale, continuously steered to the atomic fountains, even more regular calibrations of TAI, improved measurements of optical frequencies, enhanced tests of fundamental physics, measurement of the gravitational red shift within the ACES project, and remote clock comparisons \emph{via} ACES.

\section*{Acknowledgments}

SYRTE is SYst\`emes de R\'ef\'erence Temps-Espace. SYRTE is UMR CNRS 8630 between Centre National de la Recherche Scientifique (CNRS), Universit\'e Pierre et Marie Curie (UPMC) and Observatoire de Paris.
LNE, Laboratoire National de M\'etrologie et d'Essais, is the French National Metrology Institute.
SYRTE is a member of the Institut Francilien de Recherche sur les Atomes Froids (IFRAF) and of the C'Nano network of the R\'egion \^Ile de France.
 K.~Gibble acknowledges support from the NSF, Penn State, and la
Ville de Paris. We acknowledge a tremendous number of past contributions over 2 decades of development of the LNE-SYRTE fountain ensemble. We specifically acknowledge the contribution of SYRTE technical
services. The ACES mission is an ESA project. Within ACES, the PHARAO cold atom space clock is developed and funded by CNES. The transportable fountain FOM is also funded by CNES to a large extent.

\bibliographystyle{IEEEtran}

\end{document}